\documentclass[prx,amsmath,amssymb, notitlepage, twocolumn,superscriptaddress,longbibliography]{revtex4-2}

\pdfoutput=1

\usepackage{amsmath}
\usepackage{tabularx,graphicx}
\usepackage{epstopdf}
\usepackage{graphicx}
\usepackage{latexsym}
\usepackage{amssymb}
\usepackage{amsmath}
\usepackage{color, colortbl}
\usepackage{psfrag}
\usepackage{bbm}
\usepackage{bm}
\usepackage{titlesec}
\usepackage{dsfont}
\usepackage{feynmp}
\usepackage{slashed}
\usepackage{multirow}

\usepackage[tight]{subfigure}

\usepackage[papersize={8.5in,11in}]{geometry}

\usepackage{color}
\definecolor{darkblue}{rgb}{0.,0.,0.4}
\definecolor{darkred}{rgb}{0.5,0.,0.}
\definecolor{BlueViolet}{RGB}{138,43,226}
\definecolor{SkyBlue}{RGB}{30,144,255}
\definecolor{DarkGreen}{RGB}{0,100,0}
\usepackage[pdftex,colorlinks=true,linkcolor=darkblue,citecolor=blue,urlcolor=darkred]{hyperref}

\def\be{\begin{eqnarray}}
\def\ee{\end{eqnarray}}
\def \be{\begin{equation}}
\def \ee{\end{equation}}
\def \bea{\begin{eqnarray}}
\def \eea{\end{eqnarray}}
\def \nn{\nonumber \\}

\begin{document}

\title{Transport in the non-Fermi liquid phase of isotropic Luttinger semimetals}

\author{Ipsita Mandal}
\affiliation{Faculty of Science and Technology, University of Stavanger, 4036 Stavanger, Norway}
\affiliation{Institute of Nuclear Physics, Polish Academy of Sciences, PL-31342 Krak\'{o}w, Poland}

\author{Hermann Freire}
\affiliation{Instituto de F{\'i}sica, Universidade Federal de Goi{\'a}s, 74.001-970,
Goi{\^a}nia-GO, Brazil}

\begin{abstract}
Luttinger semimetals have quadratic band crossings at the Brillouin zone-center in three spatial dimensions. Coulomb interactions in a model that describes these systems stabilize a non-trivial fixed point associated with a non-Fermi liquid state, also known as the Luttinger-Abrikosov-Beneslavskii phase. We calculate the optical conductivity $\sigma (\omega) $ and the dc conductivity $\sigma_{dc} (T) $ of this phase, by means of the Kubo formula and the Mori-Zwanzig memory matrix method, respectively. Interestingly, we find that $\sigma (\omega) $, as a function of the frequency $\omega$ of an applied ac electric field, is characterized by a small violation of the hyperscaling property in the clean limit, which is in  contrast with the low-energy effective theories that possess Dirac quasiparticles in the excitation spectrum and obey hyperscaling. Furthermore, the effects of weak short-ranged disorder on the temperature dependence of $\sigma_{dc} (T)$ give rise to a stronger power-law suppression at low temperatures compared to the clean limit. Our findings demonstrate that these disordered systems are actually power-law insulators. Our theoretical results agree qualitatively with the data from recent experiments performed on Luttinger semimetal compounds like the pyrochlore iridates [(Y$_{1-x}$Pr$_x$)$_2$Ir$_2$O$_7$].
%%%%%%%%%%%%%%
\end{abstract}

\maketitle

\tableofcontents

\section{Introduction}

Theories of non-Fermi liquid (NFL) phases in two and three-dimensions are one of the biggest enigmas in the field of strongly-correlated quantum matter and even today, after many decades of intense research, remain largely an unsolved problem. A deep understanding of these NFL phases turns out to be crucial in view of the fact that these states
naturally lead to new emergent phases (such as high-temperature superconductivity, topological phenomena in semi-metals and superconductors, etc) as some external parameter like temperature, pressure or doping is varied in the system.
It is a theoretically challenging task to study such systems, and consequently there have been intensive efforts dedicated to building a framework to understand them  \cite{nayak,nayak1,lawler1,mross,Jiang,ips2,ips3,Shouvik1,Lee-Dalid,shouvik2,Freire_Pepin_2,ips-uv-ir1,Freire_Pepin_1,ips-uv-ir2,ips-subir,ips-sc,ips-c2,Lee_2018,ips-fflo,ips-nfl-u1}. They are also referred to as critical Fermi surface states, as the breakdown of the Fermi liquid theory is brought about by the interplay between the soft fluctuations of the Fermi surface and some gapless bosonic fluctuations.

Recently, there has been also an upsurge of interest in a new frontier of this field where NFL phases can be observed at a Fermi point, i.e., in the absence of a large Fermi surface.
From the analysis of the electronic structure of compounds like pyrochlore iridates, the half-Heusler compounds, and grey-Sn, a minimal effective model to describe such systems turns out to be the well-known three-dimensional Luttinger model with quadratic band crossings at the zone-center (i.e., the $\Gamma$ point). Consequently, the materials
that are well-described by this low-energy effective theory are nowadays known as ``Luttinger semimetals'' in the literature \cite{moon-xu,rahul-sid,ips-rahul,*ips-rahul-errata,ips-qbt-sc,ips_qbt_plasmons,ips_qbt_tunnel}.
This novel class of materials not only exhibits strong spin-orbit coupling, but also has strong electron-electron interactions. Since electron-electron interactions are
not screened in these systems, an effective description must also include long-range Coulomb interactions.  Interestingly, this problem was studied for the first time back in 1974 by Abrikosov \cite{Abrikosov}, who demonstrated, using renormalization group (RG) arguments, that the Coulomb interaction in the model stabilizes a non-trivial fixed point associated with a new NFL state in three spatial dimensions, which was later called the Luttinger-Abrikosov-Beneslavskii (LAB) phase \cite{moon-xu}. This fixed point is stable provided that time-reversal symmetry and the cubic symmetries are preserved in the system. This earlier work was later rediscovered and extended by Moon \emph{et al.} \cite{moon-xu}, who calculated the universal power-law exponents describing various physical quantities in this LAB phase in the clean (i.e. disorder-free) limit, including the conductivity, susceptibility, specific heat, and the magnetic Gruneisen number.

From a strictly theoretical point of view, there has also been an increasing interest in the LAB phase, since it may realize the so-called ``minimal-viscosity" scenario \cite{Herbut-PRB}, in which the ratio of the shear viscosity $\eta$ with the entropy $s$ is close to the Kovtun-Son-Starinets ratio \cite{DTSon-PRL_2005}, i.e., $\eta/s\gtrsim 1/(4\pi)$. This means that these systems may be considered as a new example of a strongly-interacting ``nearly-perfect fluid''. Other important examples that satisfy this condition include the hydrodynamical fluid that emerges in a clean single-layer graphene sheet at charge neutrality point \cite{Fritz-PRB}, the quark-gluon plasma \cite{DTSon-PRL} generated in relativistic heavy-ion colliders, and ultracold fermionic gases tuned to the unitarity limit \cite{Cao}.

Naturally, transport properties of NFL phases are extremely important in order to characterize these systems. One of the widely used methods to calculate non-equilibrium properties  is the application of the quantum Boltzmann equation. This method has many merits, and along with the well-established $\varepsilon$-expansion, it has been successfully used to discuss the hydrodynamical regime of many quantum critical systems. However, this approach also has some limitations, as one of its main assumptions is that the quasiparticle excitations exist even at low energies in the model, which is of course not valid at the LAB fixed point. Therefore, alternative methods to calculate transport properties, which do not rely on the existence of quasiparticles at low energies, should be used instead in order to provide an unbiased evaluation of such properties in NFL systems at low temperatures. 
For this reason, in the present work, we will apply the Kubo formula, and also its implementation using the Mori-Zwanzig memory matrix formalism, to the Luttinger model with long-range Coulomb interactions, in order to describe some transport coefficients of the LAB phase. More specifically, we will compute the optical conductivity $\sigma(\omega)$ at $T=0$ as a function of the frequency $\omega$ of an applied ac electric field, and the dc resistivity $\rho(T)$ as a function of temperature $T$ with the addition of weak short-ranged disorder. Since the effects of disorder are relevant in the renormalization group flow sense
\cite{rahul-sid,ips-rahul,*ips-rahul-errata} for the LAB phase, they turn out to be important also for the study of the transport properties of the system at low temperatures.

The main results obtained in the paper are the following: We find that $\sigma(\omega)$ in the LAB phase is characterized by a small violation of the hyperscaling property in the clean limit, in contrast to the low-energy effective theories that possess Dirac quasiparticles in the excitation spectrum and obey hyperscaling.
Furthermore, on investigating the effects of weak short-ranged disorder on the dc conductivity $\sigma_{dc}(T)$, we find that $\sigma_{dc}(T)$ displays a stronger power-law suppression at low temperatures compared to the corresponding result in the clean limit. We then compare this theoretical result with the available experimental data.

The paper is structured as follows. In Sec.~\ref{model}, we define the LAB phase for the Luttinger Hamiltonian coupled with long-range Coulomb interactions. Then, we calculate the the optical conductivity of the LAB phase up to two-loop order in Sec.~\ref{sec:current}, using the Kubo formula. Next, in Sec. \ref{sec:memory_matrix}, we calculate the dc resistivity of the model as a function of temperature, with the addition of weak short-ranged disorder using the memory matrix formalism. Finally, in Sec.~\ref{end}, we end with a summary and some outlook.
Appendix~\ref{angular} illustrates the derivation of some relations involving the $\ell=2$ spherical harmonics in $d$ spatial dimensions, that are useful for the loop integrals.
The details of the two-loop calculations have been explained in Appendices~\ref{current2loop} and \ref{chi2loop}.

%%%%%%%%%%%
\section{Model}
\label{model}

%---------------------------------------
%----------- FIGURE ---------------
%---------------------------------------
\begin{figure}
\includegraphics[width=0.75\linewidth]{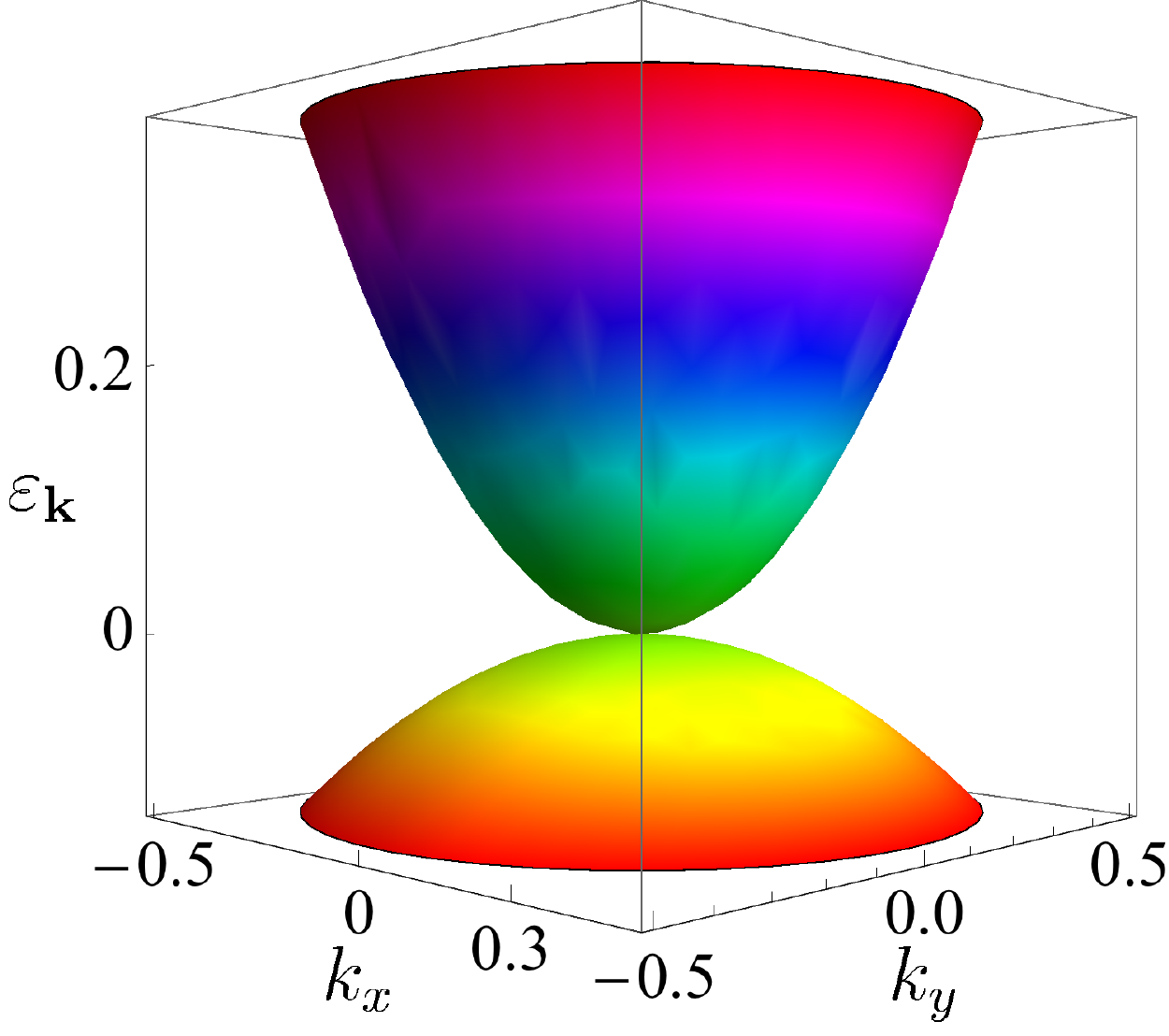}
\caption{The non-interacting dispersion $\varepsilon_{\mathbf k}$ of the isotropic Luttinger semimetal (see Eq.~\eqref{bare}) shows quadratic band-touching at the Brillouin zone-center. Here, we choose $m=1$ and $m'=0.5$. For visualization, $\varepsilon_{\mathbf k}$ is shown as a function of $k_x$ and $k_y$ (i.e., we set $k_z=0$).
\label{fig:bands}}
\end{figure}
%%%%%%%%%%%

%----------- FIGURE ---------------
%---------------------------------------
\begin{figure}
\includegraphics[width=0.65\linewidth]{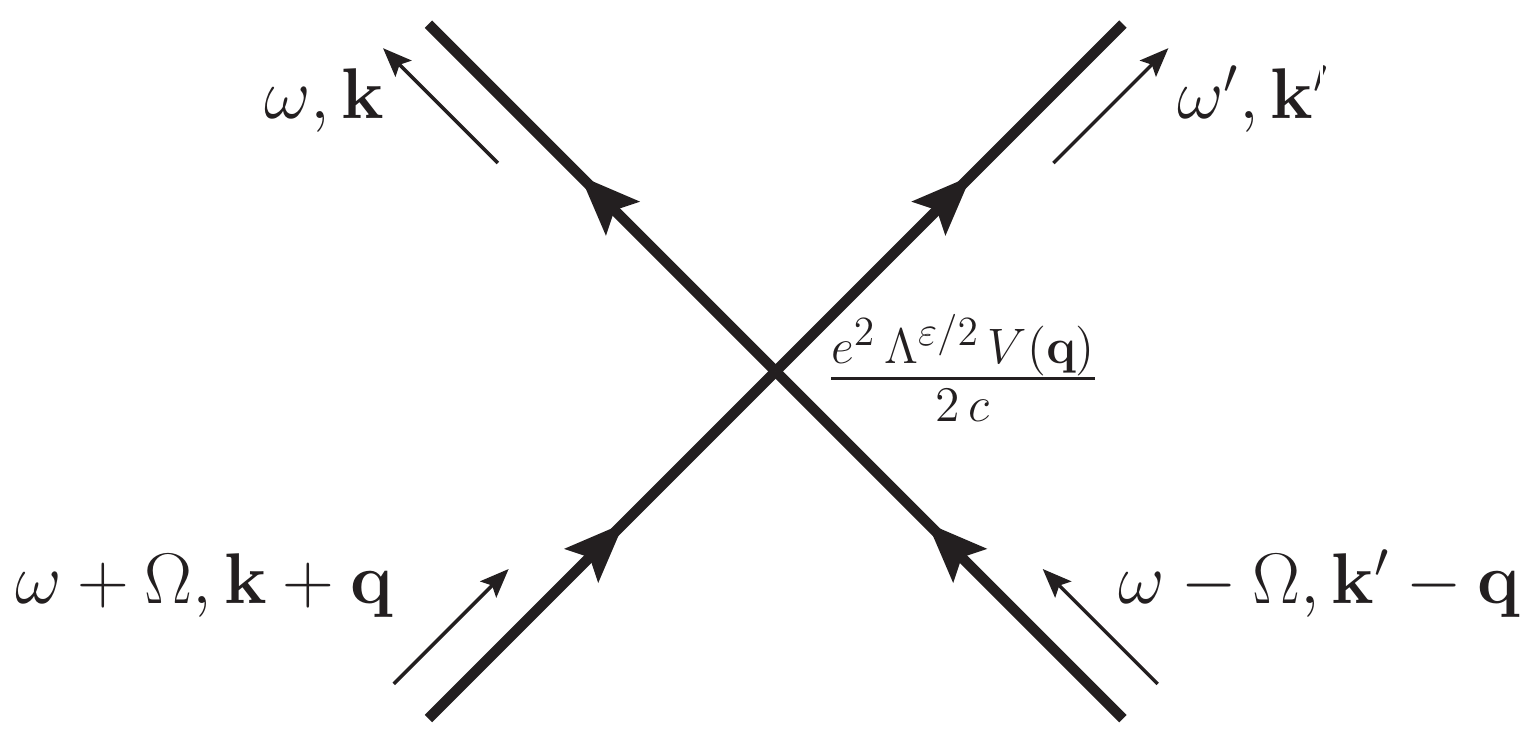}
\caption{The four-fermion vertex arising due to Coulomb interactions.
\label{fig:vert}}
\end{figure}
%%%%%%%%%%%

We consider a spin-orbit coupled system, in which the states near $\mathbf{k}=0$ at the Fermi energy are split into four-fold degenerate angular momentum $j=3/2$ states. The $\mathbf{k} \cdot \mathbf{p}$ Hamiltonian for the non-interacting system takes the following effective form:
\begin{equation}\label{bare2}
 \mathcal{H}_0 = \frac{|\mathbf k |^2}{2 \,m'}
 -\frac{\frac{5}{4} |\mathbf k |^2-(\mathbf{k}\cdot
 \boldsymbol{\mathcal J})^2}{2\,m}\,,
\end{equation}
where $\boldsymbol{\mathcal J}$ is the three-vector of the angular momentum operators transforming as the $T_2$ representation of the cubic group.
This model is also known as the Luttinger Hamiltonian \cite{Luttinger}. The system harbors quadratic band crossings at the Brillouin zone-center in three spatial dimensions (see Fig.~\ref{fig:bands}), where the low-energy bands can be cast in terms of
a four-dimensional representation of the lattice symmetry group \cite{murakami,moon-xu,igor16} as follows:
\begin{equation}\label{bare}
 \mathcal{H}_0 = \sum_{a=1}^5 d_a(\mathbf{k}) \,  \Gamma_a   
 + \frac{ |\mathbf k |^2}{2\,m'} \,,
\quad d_a  (\mathbf{k})   = \frac{\tilde d_a(\mathbf{k})   }{2\, m}\,,
 \end{equation}
%%%%%%%%%%%%%%%%
where the $\Gamma_a$ matrices are the rank-four irreducible representations of the Clifford algebra relation $\{\Gamma_a,\Gamma_b\} = 2\, \delta_{ab}$ in the Euclidean space. We have used the common notation $\{A,B\} = AB+BA$ for denoting the anticommutator. There are five such matrices that are related to the familiar gamma matrices of the Dirac equation (plus the matrix conventionally denoted as $\Gamma_5$), but with the Euclidean metric (instead of the Minkowski metric). In $d=3$, the space of $4\times 4$ Hermitian matrices is spanned by the identity matrix, the five $4\times 4$ Gamma matrices $\Gamma_a$, and the ten distinct matrices $\Gamma_{ab} = \frac{1}{2\,\mathrm{i}}\, [\Gamma_a, \Gamma_b]$.
Furthermore, the  $\tilde d_a(\mathbf k)$'s are the $\ell=2$ spherical harmonics that have the following structure:
\begin{align}
\label{ddef}
&\tilde d_1(\mathbf{k}) = \sqrt{3}\, k_y \,k_z\,,
\quad \tilde d_2(\mathbf{k}) =  \sqrt{3}\, k_x\, k_z\, ,\quad
 \tilde d_3(\mathbf{k}) =  \sqrt{3} \,k_x\, k_y\, ,\quad\nonumber\\
 &\tilde d_4(\mathbf{k}) =\frac{\sqrt{3}  \,  (k_x^2 - k_y^2) }{2}\,, \quad
 \tilde d_5(\mathbf{k}) = \frac{2\, k_z^2 - k_x^2 - k_y^2}{2} \,.
\end{align}
The isotropic $\frac{k^2}{2\,m'}$ term in Eq. \eqref{bare} with no spinor structure makes the band masses of the conduction and valence bands unequal. 

The Euclidean action of the interacting system can be written as:
\begin{align}\label{action}
S_0 =&  \int d\tau \,d^3{\mathbf x}
  \Big[ \,\sum_{i=1}^{N_f} \psi_i^{\dag}(\tau,\mathbf x)
 \left\lbrace \partial_{\tau} + \mathcal{H}_0 + \mathrm{i}\, e 
 \,\varphi(\tau,\mathbf x) \right \rbrace 
 \psi_i (\tau,\mathbf x)\nonumber\\
 & \hspace{2.2 cm}
+\frac{c}{2} \left \lbrace \nabla \varphi (\tau,\mathbf x) \right \rbrace^2  
 \Big ] \,,
\end{align}
where the Coulomb interactions are mediated by a scalar boson field $\varphi(\mathbf x)$
with no dynamics, and $N_f$ is the number of fermionic flavors (to be explained below).
%%%%%%%%%

If we integrate out the scalar boson, the Coulomb interaction shows up as an effective four-fermion term. Then the total action takes the form: 
\begin{widetext}
\begin{align}
&S = \sum_{i=1}^{N_f} 
\int \frac{d\omega \,d^3 {\mathbf k} } {(2\,\pi)^4}\, 
{\tilde \psi}_i^{\dag}(\omega,\mathbf k)
\left( -\mathrm{i} \,\omega + \mathcal{H}_0 \right)
 {\tilde \psi}_i  (\omega, \mathbf k) 
%%%%%%%%%%%%%%%%%%%%%%%
\nn & 
\qquad + \frac{e^2\,\Lambda^{\varepsilon/2}}{2 \,c } \sum_{i,i'=1}^{N_f}
\int \frac{d\omega\, d\omega' \,d\Omega\,
d^3 {\mathbf q}\, d^3 {\mathbf k}\, d^3 {\mathbf k'}}
{(2\pi)^{12}}\, 
V(|\mathbf q|)\,\tilde{\psi}^{\dag}_{i} (\omega,\mathbf k)\,
%%%%%%%%%%%%&\times 
\tilde{\psi}_{i} (\omega+\Omega,{\mathbf k}+\mathbf q) \,
\tilde{\psi}^{\dag}_{i'}( \omega' ,{\mathbf k}')\,
\tilde{\psi}_{i'}(\omega'-\Omega,{\mathbf k}'-\mathbf q)  \,,
\label{fullaction}
\end{align}
\end{widetext}
where the Coulomb interaction vertex is given by $\frac{e^2\,\Lambda^{\varepsilon/2}}{2 \,c }V(|\mathbf q|)$ (see also Fig.~\ref{fig:vert}), with $V(|\mathbf q|) = \frac{1}{{\mathbf q}^2}$,  in the momentum space. The tilde over $\psi_i$ indicates that it is the Fourier-transformed version.
We have also scaled $ e^2 $ by using the floating mass scale $\Lambda^2$ (of the renormalization group flow) to make it dimensionless for $d=4-\varepsilon $ spatial dimensions, after setting the tree-level scaling mass dimension $[\mathbf k]$ of $\mathbf k$ as unity.

The bare Green's function for each fermionic flavor is given by
\begin{align}
G_0(\omega, \mathbf{k}) =  
\frac{ \mathrm{i}\, \omega- \frac{ \mathbf k^2}{2\,m'}  + \mathbf{d}(\mathbf{k}) \cdot{\mathbf{\Gamma}}}
{-\left ( \mathrm{i}\, \omega- \frac{\mathbf k ^2}{2\, m'}  \right )^2 +|\mathbf{d}(\mathbf{k})|^2}\,,
\label{baregf}
\end{align}
where $|\mathbf{d}(\mathbf{k})|^2 = \frac{ \mathbf k^4} {4\,m^2}$. On occasions, to lighten the notation, we will use $\mathbf{d}_{\mathbf{k}}$ to denote $\mathbf{d}(\mathbf{k})$. 

This system turns out to be an NFL, which can be analyzed by a controlled approximation using dimensional regularization \cite{Abrikosov,moon-xu}.
The LAB fixed point for the clean system is given by $e=e^*$, where
\begin{align}
{e^*}^2= \frac{60\,\pi^2\,c\,\varepsilon} 
{m\,\left(4 +15\,N_f\right)},
\end{align}
and the dynamical critical exponent $z$ at this fixed point is given by $z^*=2-4\,\varepsilon/(4+15\,N_f)$ \cite{moon-xu}, where $\varepsilon=4-d$, with $d$ being the number of spatial dimensions. It is to be noted that the results obtained using dimensional regularization can also be obtained by large-$N_f$ methods. Hence, we have considered here a setting with $N_f$ independent fermionic flavors, although the physical case corresponds to $N_f=1$.

Using the Noether's theorem (see, e.g., Ref.~\cite{Peskin}), the current $\mathbf{J}$ and momentum $\mathbf{P}$ operators of the Luttinger semimetal are given by:
%\begin{align}
%\mathbf{J}&=\sum_{i}\int\frac{d^d\mathbf{k}}{(2\,\pi)^3}\,
%\tilde{\psi}_i^{\dagger}(\mathbf{k})
%\left [\nabla_{\mathbf{k}}\mathbf{d(k)}\cdot\mathbf{\Gamma} \right ]
%\tilde{\psi}_i(\mathbf{k}),\quad\nonumber\\
%\mathbf{P} &=\sum_{i}\int\frac{d^d\mathbf{k}}{(2\pi)^3}
%\, \mathbf{k}\,\tilde{\psi}_i^{\dagger}(\mathbf{k})\,\tilde{\psi}_i(\mathbf{k})\,,
%\end{align}
\begin{align}
& \mathbf{J}(q_0,\mathbf q)
\nn &=\sum_{i}\int\frac{ dk_0 \,d^d\mathbf{k}}
{(2\,\pi)^{d+1}}\,
\tilde{\psi}_i^{\dagger}(k_0+q_0, \mathbf{k} +\mathbf q)
\left [\nabla_{\mathbf{k}}\mathbf{d(k)}\cdot\mathbf{\Gamma} \right ]
\tilde{\psi}_i(k_0,\mathbf{k})\,,
\nn
& \mathbf{P} (q_0,\mathbf{q}) \nn
&=\sum_{i} \int\frac{ dk_0 \,d^d\mathbf{k}}
{(2\,\pi)^{d+1}}\,
\left(  \mathbf{k} +\mathbf{q}/2 \right)
\tilde{\psi}_i^{\dagger}(k_0+q_0,\mathbf{k}+\mathbf q)\,
\tilde{\psi}_i(k_0,\mathbf{k})\,,
\end{align}
which are associated with the global U(1) symmetry and continuous spatial translation invariance, respectively, of Eq.~\eqref{action}.
%%%%%%%%%%%
In the rest of the paper, we will consider the case with $N_f=1$.

%%%%%%%%%%%%%%%%%%%%%%%%%%%%%%%%%%%%%%%%
\section{Current-current correlation function and optical conductivity}
\label{sec:current}

In this section, we will compute the optical conductivity $\sigma(\omega) = 
\sigma_{zz}(\omega, \mathbf{q} = \mathbf{0})$ at $T = 0$ via the Kubo formula
\begin{equation}
	\sigma(\omega) = -\frac{\left \langle J_z \,J_z \right \rangle(\mathrm{i}\,\Omega)}{\Omega} 
\bigg|_{\mathrm{i}\, \Omega \rightarrow \omega + \mathrm{i}\,0^+}\,,
\label{eq:Kubo}
\end{equation}
for current flowing along the $z$-direction. Here we will consider the case with equal band masses, i.e., $m' =\infty $.
Since the model is isotropic, the scaling relation is not dependent on the choice of the direction of the current flow.
We take an approach similar to the ones taken in the context of NFL models in the presence of a large Fermi surface \cite{subir-aavishkar,ips-subir,ips-c2}.

%%%%%%%%%%%%%%%%%%%%%%%%%%%%
\begin{figure}[t]
	\centering
	\includegraphics[width=0.3\textwidth]{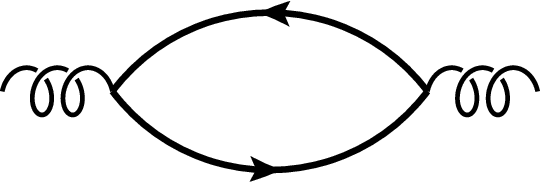} \label{fd1}
	\caption{Feynman diagram for the contribution to the current-current 
correlation function at one-loop order.}
	\label{fig1loop}
\end{figure}
%%%%%%%%%%%%%%%%%%%%%%%%%%%%%%%%%%%%%%%%%5

We will employ the scheme developed by Moon {\it et al.} \cite{moon-xu}, where the radial momentum integrals are performed with respect to a $d=4-\varepsilon$ dimensional measure $\int 
\frac{ |\mathbf k|^{3-\varepsilon} d|\mathbf k| }{(2\,\pi)^{4-\varepsilon}}$, but the $\Gamma$ matrix structure is as in $d=3$. The angular integrals are performed only over the three-dimensional sphere parameterized by the polar and azimuthal angles $(\theta, \,\phi)$. However, the overall angular integral of an isotropic function $\int_{\hat{\Omega}}\cdot 1$ is taken to be $2 \,\pi^2$ (which is appropriate for the total solid angle in $d=4$), and the angular integrals are normalized accordingly. Therefore, the angular integrations are performed with respect to the following measure: 
\begin{equation}
\label{moonmeasure}
\int dS\, (\ldots) \equiv \frac{\pi}{2} \int_0^{\pi} d \theta \int_0^{2\pi} 
d \phi\, \sin \theta \, (\ldots)\,,
\end{equation}
where the $\pi/2$ is inserted for the sake of normalization.
To perform the full loop integrals, we will use the relations shown in Appendix~\ref{angular}.

%%%%%%%%%%%%%%%%%%%%%%%%
\subsection{One-loop contribution}

The current-current correlation function at 
one-loop level (see Refs.~\cite{Broerman_1,Broerman_2,Boettcher_PRB_2019,Witczak-Krempa_PRB_2019,Polini_PRB_2019} for related work) is given by a simple fermionic 
loop with two current insertions, as shown in Fig.~\ref{fig1loop}. In the present model, it evaluates to
\begin{widetext}
\begin{align}
& \langle J_z J_z \rangle_\text{1loop}(\mathrm{i}\,\omega)
= -  
\int \frac{dk_0}{2\pi} \int \frac {d^d\mathbf{k}} {(2\pi)^d} 
\text{Tr} \left [ \left \lbrace 
\partial_{k_z}\mathbf{d}(\mathbf k )\cdot \mathbf \Gamma\right \rbrace
G_0(k+q) \left \lbrace 
\partial_{k_z}\mathbf{d}(\mathbf k)\cdot \mathbf \Gamma\right \rbrace G_0(k)\right ]\nn
%%%%%%%%%%%%%%%%%
&=  -  \int \frac{dk_0}{2\pi} \int \frac {d^d\mathbf{k}} {(2\pi)^d}  
\text{Tr} \left [ \left \lbrace 
\partial_{k_z}\mathbf{d}(\mathbf k)\cdot \mathbf \Gamma\right \rbrace
\frac{ \mathrm{i}\, k_0 +\mathrm{i}\,\omega + \mathbf{d}(\mathbf{k} ) \cdot{\mathbf{\Gamma}}}
{-\left ( \mathrm{i}\, k_0 +\mathrm{i}\,\omega \right )^2 +|\mathbf{d}(\mathbf{k})|^2}\,
 \left \lbrace 
\partial_{k_z}\mathbf{d}(\mathbf k)\cdot \mathbf  \Gamma\right \rbrace
\frac{ \mathrm{i}\, k_0  + \mathbf{d}(\mathbf{k}) \cdot{\mathbf{\Gamma}}}
{-\left ( \mathrm{i}\, k_0   \right )^2 +|\mathbf{d}(\mathbf{k})|^2}
\right ]\nn
%%%%%%%%%%%%%%%
&= - 4 
  \int \frac{dk_0}{2\pi} \int \frac {d^d\mathbf{k}} {(2\pi)^d}  
\frac{ - \left \lbrace 
\partial_{k_z}\mathbf{d}(\mathbf k)\right \rbrace^2\,( k_0 +\omega)\,k_0 
+  \frac{1}{4} \left \lbrace 
\partial_{k_z}  \mathbf{d}^2(\mathbf k)\right \rbrace^ 2
-  \left \lbrace 
\partial_{k_z}\mathbf{d}(\mathbf k)\right \rbrace^2 \mathbf{d}^2(\mathbf k) }
%%%
{ \left [ -\left ( \mathrm{i}\, k_0 +\mathrm{i}\,\omega \right )^2 +|\mathbf{d}(\mathbf{k})|^2
\right ] 
\left [ -\left ( \mathrm{i}\, k_0   \right )^2 +|\mathbf{d}(\mathbf{k})|^2 \right] }\nn
%%%%%%%%%%%%%%%%%%%%%%%%%
& = - 
\frac{ m^{1-\frac{\varepsilon }{2}} | \omega | ^{2-\frac{\varepsilon }{2}}}
{\pi ^2 \,\varepsilon }
\,,
%%%%%%%%%%%%%%%
\label{eq:jj_1Loop}
\end{align}
\end{widetext}
where $q=(\omega,\,\mathbf 0)$.
Consequently, at zeroth order, the optical conductivity $\sigma(\omega)$ is proportional to $\omega^{1-\frac{\varepsilon }{2}}$. In $d=4-\varepsilon $, this result
then agrees with the so-called hyperscaling property, where the optical conductivity is expected to scale as $\sigma(\omega)\sim\omega^{(d-2)/z}$ for $\omega\gg T$.

In the next subsection, we will consider the effect of the Coulomb interactions, and show how this affects the hyperscaling property of the Luttinger semimetal.

%%%%%%%%%%%%%%%%%%%%%%%%%%%%%%%%%%%%%%%%%%%%%%%%%%%%

%%%%%%%%%%%%%%%%%%%%%%%%
\subsection{Two-loop contributions}

%%%%%%%%%%%%%%%%%%%%%%%%%%%%
\begin{figure*}[t]
	\centering
	\subfigure[]{\includegraphics[width=0.3\textwidth]{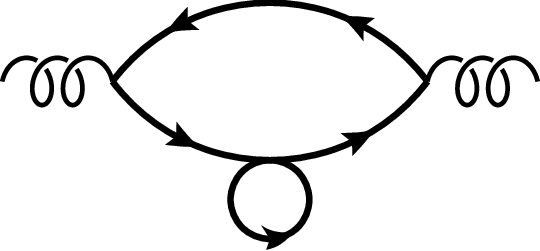} \label{fig2}}
	\subfigure[]{\includegraphics[width=0.3\textwidth]{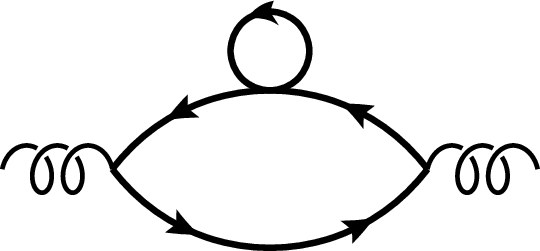} \label{fig3}}
	\subfigure[]{\includegraphics[width=0.3\textwidth]{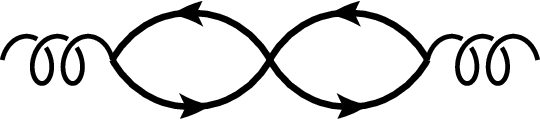} \label{fig4}}
	\caption{\label{fig2loop}Feynman diagrams for the contributions to the current-current 
correlation function at two-loop order. (a) and (b) represent the diagrams with self-energy corrections, while (c) corresponds to the diagram with vertex correction.}
\end{figure*}

At two loops, we obtain three Feynman diagrams, as shown in Figs.~\ref{fig2}, \ref{fig3}, and \ref{fig4}. The first two diagrams (Figs.~\ref{fig2} and \ref{fig3}) correspond to the fermion self-energy corrections (due to the Coulomb interactions), given by the insertion of the one-loop rainbow graph to the current-current correlator.
We include a factor of $2$, since the diagrams in Figs.~\ref{fig2} and \ref{fig3} give equal contributions. This yields the result
\begin{align}
\langle J_z J_z \rangle_\text{2loop}^{(1)}(\mathrm{i}\,\omega) 
&=  
\frac{e^2 \, m^{2-\frac{\varepsilon }{2}} \,| \omega | ^{2-\frac{\varepsilon }{2}}}
{90\, \pi ^4 \,c \,\varepsilon ^2} 
\left( \frac{\Lambda}{m\,|\omega|} \right)^{\varepsilon/2}\nonumber\\
&- \frac{e^2 \, m^{2-\frac{\varepsilon }{2}} | \omega | ^{2-\frac{\varepsilon }{2}} 
\ln\left(\frac{m \,| \omega | }{\Lambda }\right)}
{180\, \pi ^4 \,c\, \varepsilon }\,.
 \end{align}
The calculational details of the above equation can be found in Appendix~\ref{app2loop1}. 
From the results presented in that Appendix, we observe that since the fermionic self-energy at one-loop level [i.e., $\Sigma_1(k) \equiv \Sigma_1(\mathbf k)$] does not have a frequency dependence, the quasiparticle weight, defined by $Z_F\,\mathcal{I}_{4\times 4} =
\left[ \mathcal{I}_{4\times 4}-\lim \limits_{k_0 \rightarrow 0} 
\lim \limits_{ \mathbf k \rightarrow 0} 
\frac{\partial}{\partial k_0} \Sigma_1(\mathbf k)\right]^{-1}$, is equal to unity at this order (but of course nonzero corrections to $Z_F$ can appear in higher-loop contributions).
Howover, if we calculate the renormalized mass $m^*$, which is given by
the standard definition
$$ \frac{m}{m^*} \,\mathbf{d}(\mathbf k) \cdot \mathbf \Gamma
= Z_F \left[ \mathbf{d}(\mathbf k) \cdot \mathbf \Gamma 
+ d_a (\mathbf k) \,\lim \limits_{k_0 \rightarrow 0} 
\lim \limits_{ \mathbf k \rightarrow 0} 
\frac{\partial}
{\partial d_a } \Sigma_1(\mathbf k)
 \right],$$ we obtain $ m^* \rightarrow 0$.
%However, the renormalized mass $m^{*}$ is given by
%$$ \frac{m}{m^*}= 1
%- \frac{\partial \Sigma_1(\mathbf k)}
%{\partial(\mathbf d_{\mathbf k} \cdot \Gamma)}
%$$, which gives $m^{*}\rightarrow \infty$ as $\varepsilon \rightarrow 0$.

As for the diagram in Fig.~\ref{fig4}, which refers to the simplest vertex correction, it evaluates to
\begin{align}
\langle J_z J_z \rangle_\text{2loop}^{(2)}(\mathrm{i}\,\omega) 
&= \frac{ e^2\,  
m^{2-\frac{\varepsilon }{2}} \,| \omega | ^{2-\frac{\varepsilon }{2}}
\left( \frac{\Lambda}{m\,|\omega|} \right)^{\varepsilon/2}
}
{ 60 \,\pi ^4 \,c\, \varepsilon ^2}\nonumber\\
&-\frac{  e^2 \, m^{2-\frac{\varepsilon }{2}} \,| \omega | ^{2-\frac{\varepsilon }{2}} \ln\left(\frac{m | \omega | }{\Lambda }\right)}
{ 120 \,\pi ^4\, c \,\varepsilon }\,.
\end{align} 
Note that this vertex corresponds to the four-fermion vertex (see Fig.~\ref{fig:vert}), which arises from Coulomb interactions. Again, the details of the calculations can be found in Appendix~\ref{app2loop2}.

%%%%%%%%%%%%%%%%%%%%%%%%%%%%%%%%%%%%%
\subsection{Scaling of the optical conductivity up to two-loop order}

In order to obtain the renormalized quantity in the effective field theory model, we have to use the fact that $\frac{1} {\varepsilon^2}$ terms are cancelled by the corresponding counterterms of the renormalized action \cite{Peskin}. We also use the value $\frac{m\,{e^*}^2}
{ \pi^2\, c}= \frac{ 60\,\varepsilon} 
{ 19 }$ at the NFL fixed point. Gathering all the terms, the final expression for $\langle J_z J_z \rangle$ up to two-loop order takes the form:
\begin{widetext}
\begin{align}
\langle J_z J_z \rangle (\mathrm{i}\,\omega)
& = \langle J_z J_z \rangle_\text{1loop}(\mathrm{i}\,\omega)
+ \langle J_z J_z \rangle_\text{2loop}^{(1)}(\mathrm{i}\,\omega) 
+\langle J_z J_z \rangle_\text{2loop}^{(2)}(\mathrm{i}\,\omega) 
+\langle J_z J_z \rangle_\text{counterterms}^{(1)}(\mathrm{i}\,\omega) 
\nn &
=  - 
\frac{ m^{1-\frac{\varepsilon }{2}} | \omega | ^{2-\frac{\varepsilon }{2}}}
{\pi ^2 \,\varepsilon }
- \frac{ {e^*}^2 \, m^{2-\frac{\varepsilon }{2}} | \omega | ^{2-\frac{\varepsilon }{2}} 
\ln\left(\frac{m | \omega | }{\Lambda }\right)}
{180\, \pi ^4 \,c\, \varepsilon }
-\frac{  {e^*}^2 \,m^{2-\frac{\varepsilon }{2}} \,| \omega | ^{2-\frac{\varepsilon }{2}} \ln\left(\frac{m \,| \omega | }{\Lambda }\right)}
{ 120\, \pi ^4\, c \,\varepsilon }
%%%%%%%%%%%%%%%%%%%%%%%%%%%%%%
\nn &
=  - 
\frac{ m^{1-\frac{\varepsilon }{2}} | \omega | ^{2-\frac{\varepsilon }{2}}}
{\pi ^2 \,\varepsilon }
\left[ 1 +\frac{ 5 \,\varepsilon  }
{ 114 }
\ln\left(\frac{m\, | \omega |}  {\Lambda }  \right)
 \right] \nn
%%%%%%%%%%%%%%%%%%%% 
& \simeq - \frac{m^{1-\frac{\varepsilon }{2}} 
\,| \omega | ^{2-\frac{\varepsilon }{2}
+ \frac{ 5 \,\varepsilon  }{ 114}  
}}
{\pi ^2 \,\varepsilon }
\left( \frac{m }{\Lambda }\right)
^{\frac{5 \,\varepsilon } { 114 }
}\,,
\end{align}
\end{widetext}
after re-exponentiating the correction term coming from the two-loop diagrams.
Therefore, the corrected optical conductivity scales as 
\begin{equation}
\sigma(\omega)\sim \omega^{1-\frac{\varepsilon }{2}
+ \frac{ 5 \,\varepsilon } {114}  }\,,
\end{equation}
after including the leading order corrections. 

Since the optical conductivity does not scale as $ \omega^{(d-2)/z^*}$, where $z^*$ is the dynamical critical exponent at the LAB fixed point, we conclude that there exists a small violation (proportional to $\varepsilon$) of the hyperscaling for the optical conductivity in the LAB phase. This should be contrasted with other effective theories that possess Dirac quasiparticles in the excitation spectrum, and obey hyperscaling.

%%%%%%%%%%%%%%%%%%%%%%%%%%%%%%%%%%%%%%%%%%%
\section{Memory matrix formalism}
\label{sec:memory_matrix}

The second method that we will use in this work to calculate transport properties is the Mori-Zwanzig memory matrix approach (see Refs.~\cite{Forster-HFBSCF,Rosch-PRL,Hartnoll-PRB_2013,Freire-AP_2014,Patel-PRB,Hartnoll_PRB_2014,Zaanen-CUP,Freire-AP_2017,Freire-EPL,Sachdev-MIT,Freire-EPL_2018,Berg-PRB,Freire-AP_2020,wang2020low}, for many successful applications of this formalism in various recent works). This method turns out to be ideal to describe the strongly interacting regime of the LAB phase, since: (1) it is not based on the existence of well-defined quasiparticles at low energies, and (2) it can correctly describe the effective nearly-hydrodynamic regime that is expected to govern the complicated non-equilibrium dynamics of these systems. Here, we will be concise in explaining the technicalities of this formalism, as more details can be found in the literature \cite{Freire-AP_2017,Freire-AP_2020}. In this framework, the matrix of conductivities can be written as: 
\begin{align}
\sigma(\omega,T)=\frac{\chi^R_{JP}(T)}
{\left  [ M_{PP}(T)-\mathrm{i} \,\omega \,\chi^R_{JP}(T)  \right ]
\left [\chi^R_{JP}(T)\right  ]^{-1}}\,,
\end{align}
with $\chi^R_{JP}(T)$ being the static retarded susceptibility (which gives the overlap of the current and momentum in the model), and $M_{PP}(T)$ is the memory matrix. For transport along the $z$-direction, $\chi^R_{JP}(T)$ is given by:
\begin{align}
\chi_{J_zP_z}(T)=\int_{0}^{\beta}d\tau 
\left \langle J_z(\tau)\, P_z(0) \right \rangle.
\end{align}
As for the memory matrix, to leading order, it is given by (again, for transport along the $z$-direction):
\begin{align}
M_{P_z P_z}(T) =\int_{0}^{\beta} d\tau 
\left \langle  \dot{P}^{\dagger}_z(0)\,
\frac{\mathrm{i}}
{\omega-L_0}\, \dot{P}_z(\mathrm{i}\,\tau) \right \rangle \,,
\end{align}
where $L_0$ is the non-interacting Liouville operator. Consequently, the dc conductivity (i.e. $\sigma_{dc}(T)\equiv\sigma(\omega\rightarrow 0,T)$) is given by
\begin{align}
\sigma_{dc}(T)=\frac{\chi^2_{J_z P_z}(T)}
{  \lim \limits _{ \omega \rightarrow 0}
\frac{\text{Im}\,G^R_{\dot{P}_{z} \dot{P}_{z}}(\omega,T)}
{\omega} }\,,
\end{align}
where $G^R_{\dot{P}_{z} \,\dot{P}_{z}}(\omega,T)
=\left \langle \dot{P}_{z}(\omega) \,
\dot{P}_{z}(-\omega)\right \rangle_0$ is the corresponding retarded correlation function in the Matsubara formalism. The notation $\langle \ldots\rangle_0$ indicates that the average is in a grand-canonical ensemble to be taken with the non-interacting Hamiltonian of the system.

One important mechanism for momentum relaxation that causes dissipation in the present transport theory is the coupling of the fermions to (weak) disorder. For this reason, we now add an impurity term that couples to the fermionic density as represented by the action:
\begin{align}
S_{imp}=\sum_{i}\int d\tau \,d^3\mathbf{x}\, W(\mathbf{x})
\, \psi_i^{\dagger}(\tau, \mathbf{x})\,\psi_i(\tau, \mathbf{x})\,.
\end{align}
We consider a weak uncorrelated disorder following a Gaussian distribution: $ \langle W(\mathbf{x})\rangle_{avg}=0$ and $ \langle W(\mathbf{x})\,W(\mathbf{x'})\rangle
=W_0\, \delta^3(\mathbf{x}-\mathbf{x'})$, where $W_0$ represents the average magnitude square of the random potential experienced by the fermionic field. Therefore, to leading order in the impurity coupling strength,
we obtain the expression:
\begin{align}
\lim_{\omega\rightarrow 0}\frac{\text{Im}\,G^{R}_{\dot{P}_{z} \dot{P}_{z}}(\omega,T)}{\omega}\approx\lim_{\omega\rightarrow 0}
W_0 \int\frac{d^3\mathbf{q}}{(2\pi)^3} \, \frac{\text{Im}\,\Pi_0^R(\mathbf{q},\omega)}{\omega}
\,,
\end{align}
where $\Pi_0^R(\mathbf{q},\omega)=\Pi_0(\mathbf{q},\mathrm{i} \,\omega\rightarrow \omega+
\mathrm{i}\, 0^+)$ is the corresponding retarded correlation function in the model, with the polarizability $\Pi_0(\mathbf{q},\mathrm{i} \,\omega)$ being given by:
\begin{align}
&\Pi_0(\mathbf{q},\mathrm{i} \,\omega)\nonumber \\
&=-T\,
\sum_{k_0}\int\frac{d^3\mathbf{k}}{(2\pi)^3}\, k_z^2
\,\text{Tr}
\big[G_0(\mathbf{k}+\mathbf{q},\mathrm{i} \,k_0+\mathrm{i} \,\omega)
\, G_0(\mathbf{k},\mathrm{i} \,k_0)\big]\,.
\end{align}

We now proceed to calculate $\chi_{J_z P_z}(T)$  and $M_{P_z P_z}(T)$ in the static limit at finite temperatures in the following subsections.
Note that, unlike in the previous section, instead of performing a systematic $\varepsilon$-expansion, we will work directly in $d=3$ to overcome technical complexity. Furthermore, we will use a hard ultraviolet (UV) cutoff $\Lambda_0$ for the the momentum integrals, rather than using a dimensional regularization.

%%%%%%%%%%%%%%%%%%%%%%%%%%%%%%%%%%%%%%%%%
\subsection{Current-momentum susceptibility at finite $T$}

First we note that for equal band masses, implemented by taking the limit $m' \rightarrow \infty $, the current-momentum susceptibility clearly vanishes at one-loop order, as only an odd power of $k_0$ appears in the numerator.
Furthermore, at two-loop order, the contribution to the current-momentum susceptibility due to self-energy insertions (similar to the diagrams depicted in Figs.~\ref{fig2} and \ref{fig3}) is given by
\begin{widetext}
\begin{align}
& -\chi_{J_z P_z}
= 
  \left(\frac{e^2}{2\,c} \right)
T^2 \sum \limits_{k_0,\ell_0}
\int \frac{d^3 {\mathbf k} \, d^3 {\boldsymbol \ell}}{(2\pi)^6} \, k_z \,
\text{Tr}
\left [ \left(\partial_{k_z}
\mathbf{d}(\mathbf{k}) \cdot \mathbf{\Gamma} \right) 
G_0(k_0, \mathbf{k})
\frac{G_0( \ell_0,  \mathbf{k} + \boldsymbol{\ell})} {\boldsymbol {\ell}^2}
G_0(k_0, \mathbf{k} )
\,G_0(k_0, \mathbf{k}) \right] \nn
%%%%%%%%%%%%%%
& =   \left(\frac{e^2}{2\,c} \right)
T^2\sum \limits_{k_0,\ell_0}
\int \frac{d^3 {\mathbf k}\, d^3 {\boldsymbol \ell}}{(2\pi)^6}\, k_z \,
\frac{ \text{Tr}
\left [ \left(\partial_{k_z}
\mathbf{d}(\mathbf{k}) \cdot \mathbf{\Gamma} \right)   
\frac{ \mathrm{i}\, k_0+ \mathbf{d}(\mathbf{k}) \cdot{\mathbf{\Gamma}}}   
{\left ( \mathrm{i}\,k_0  \right )^2
-| \mathbf{d}(\mathbf{k})|^2} 
\, \frac{  \mathbf{d}( \mathbf{k} ) \cdot{\mathbf{\Gamma}}} 
 {
\left ( \mathrm{i}\,\ell_0  \right )^2
- | \mathbf{d}( \mathbf{k} )|^2}
\, 
\frac{ \mathrm{i}\, k_0+ \mathbf{d}(\mathbf{k}) \cdot{\mathbf{\Gamma}}}   
{\left ( \mathrm{i}\,k_0  \right )^2
-| \mathbf{d}(\mathbf{k})|^2}
\,\frac{ \mathrm{i}\, k_0+ \mathbf{d}(\mathbf{k}) \cdot{\mathbf{\Gamma}}}   
{\left ( \mathrm{i}\,k_0  \right )^2
-| \mathbf{d}(\mathbf{k} )|^2} \right] }
{ \left (\mathbf k+\boldsymbol {\ell} \right ) ^2}\nn
&= 0
\,,
\end{align}
\end{widetext}
which also vanishes, as it also contains only odd powers of $k_0$ in the numerator after performing the trace in the above integral. One can verify that the same result holds for the two-loop diagram with the vertex correction, similar to Fig.~\ref{fig4}. In fact, this vanishing result holds for all higher-order loops. This is related to the particle-hole symmetry of the model, which is present for equal band masses. Since $\mathbf J$ and  $\mathbf P$ are odd and even, respectively, under particle-hole symmetry, their overlap (i.e., the current-momentum susceptibility) must be zero at all loop orders.

The vanishing of $\chi_{J_zP_z}(T)$ no longer holds for finite $m'$ (i.e., for unequal conduction and valence band masses). For this reason, we will analyze the effect of higher-order corrections of the current-momentum susceptibility for finite $m'$.

We first calculate the free fermion susceptibility. 
It evaluates to
\begin{widetext}
\begin{align}\label{chiJP}
\chi^{1loop}_{J_z P_z}(T)
& =-\lim_{\mathbf{q} \rightarrow 0} T\,
\sum \limits_{k_0}
\int \frac{d^3 {\mathbf k}}{(2\pi)^3} \,k_z \,\text{Tr}
\left [ \left(\partial_{k_z}
\mathbf{d}(\mathbf{k}) \cdot \mathbf{\Gamma} \right)  G_0(k_0, \mathbf{k}+\mathbf{q})\nonumber
\, G_0(k_0, \mathbf{k}) \right] \nn
%%%%%%%%%%%%%%%%%%%%%%%%%%%%%
&=-4\,T \, \lim_{\mathbf{q} \rightarrow 0}
\sum_{k_0}\int\frac{d^3\mathbf{k}}{(2\pi)^3} \,k_z 
\, \frac{ \left(\mathrm{i} \,k_0-\frac{(\mathbf{k+q})^2}{2m'}\right)
\frac{k_z \,k^2}{2 m^2}
+\left(\mathrm{i} \,k_0-\frac{\mathbf{k}^2}{2m'}\right)\,\partial_{k_z}\mathbf{d(k)}\cdot\mathbf{d(k+q)}
}
{\left[\left(\mathrm{i} \,k_0+\mathrm{i} \,\omega-\frac{(\mathbf{k+q})^2}{2m'}\right)^2-|\mathbf{d}_{\mathbf{k+q}}|^2\right]\left[\left(\mathrm{i} \,k_0-\frac{\mathbf{k}^2}{2m'}\right)^2-|\mathbf{d}_{\mathbf{k}}|^2\right]}.
\end{align}
\end{widetext}

\begin{figure}[]
\includegraphics[width=0.44\textwidth]{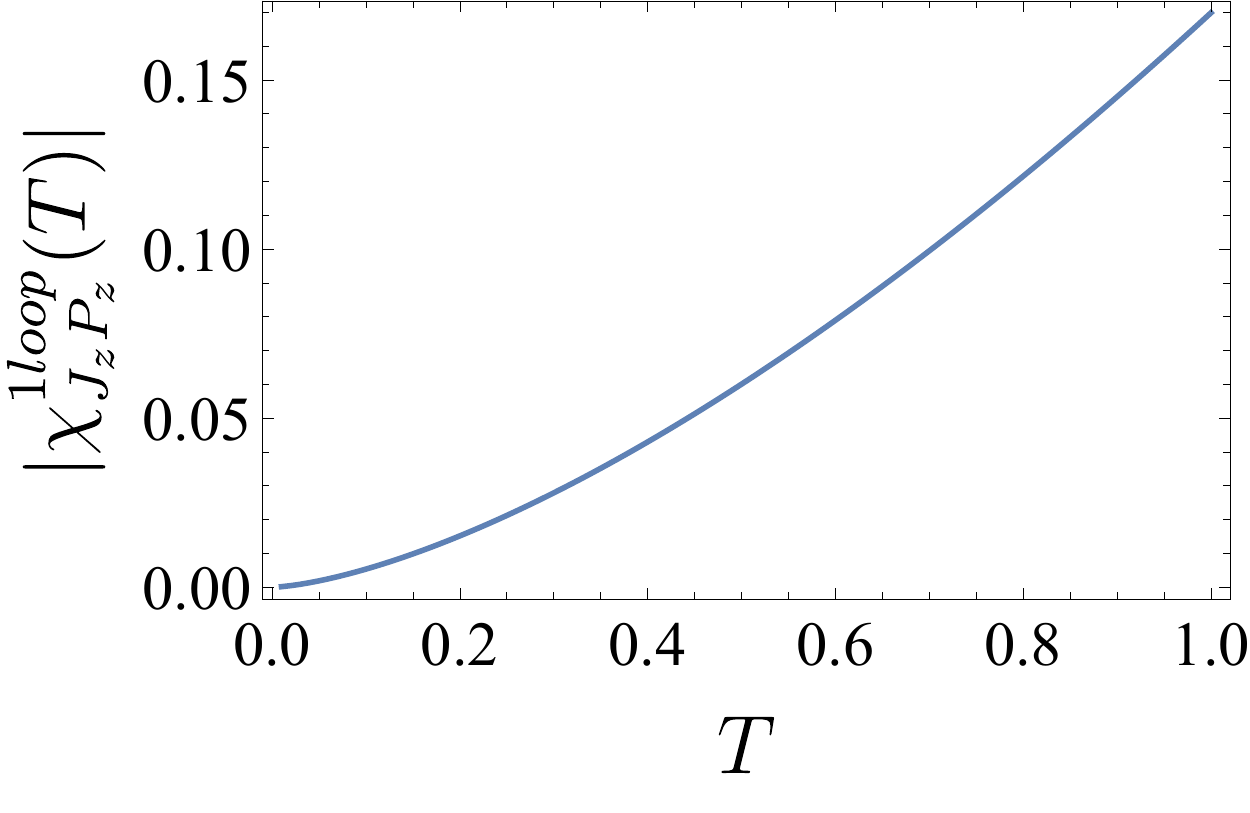}
\caption{\label{Fig:chi_JP}
Plot of the current-momentum susceptibility $\chi^{1loop}_{J_z P_z}(T)$ at one-loop order versus temperature $T$. Here, we have chosen the parameters $m=1$, $m'=5$, $N_f=1$, and the UV cutoff $\Lambda_0$ for the momentum integrals has been taken to the infinity limit (note that this result does not depend on the UV cutoff). The temperature dependence of this one-loop contribution is found to be $|\chi^{1loop}_{J_z P_z}(T)|\sim 0.170\,T^{3/2}$.} 
\end{figure}

We then perform the above summation over the fermionic Matsubara frequency $k_0$ using the method of residues using the standard formula
\begin{align}
T\sum_{\omega_n}h(\omega_n)=\sum_{\mathrm{z}_k}\text{Res}
\left [n_F( \mathrm{z} ) \,h(-\mathrm{i} \,\mathrm{z}) \right ]
\bigg |_{ \mathrm{z}_k = \text{ Poles of }h(-\mathrm{i} \,\mathrm{z})}\,,
\end{align}
where Res[\ldots] denotes the residue, and $n_F(\mathrm{z})=
\frac{1}{e^{\mathrm{z}/T}+1 }$ is the Fermi-Dirac distribution function. Next we solve Eq.~\eqref{chiJP} by means of both analytical and numerical techniques using the software {\tt Mathematica}, and obtain that $\chi^{1loop}_{J_z P_z}(T)\sim T^{3/2}$ (see Fig.~\ref{Fig:chi_JP}).

%%%%%%%%%%%%%%%%%%%%%%%%%%

\begin{figure}[]
\includegraphics[width=0.45\textwidth]{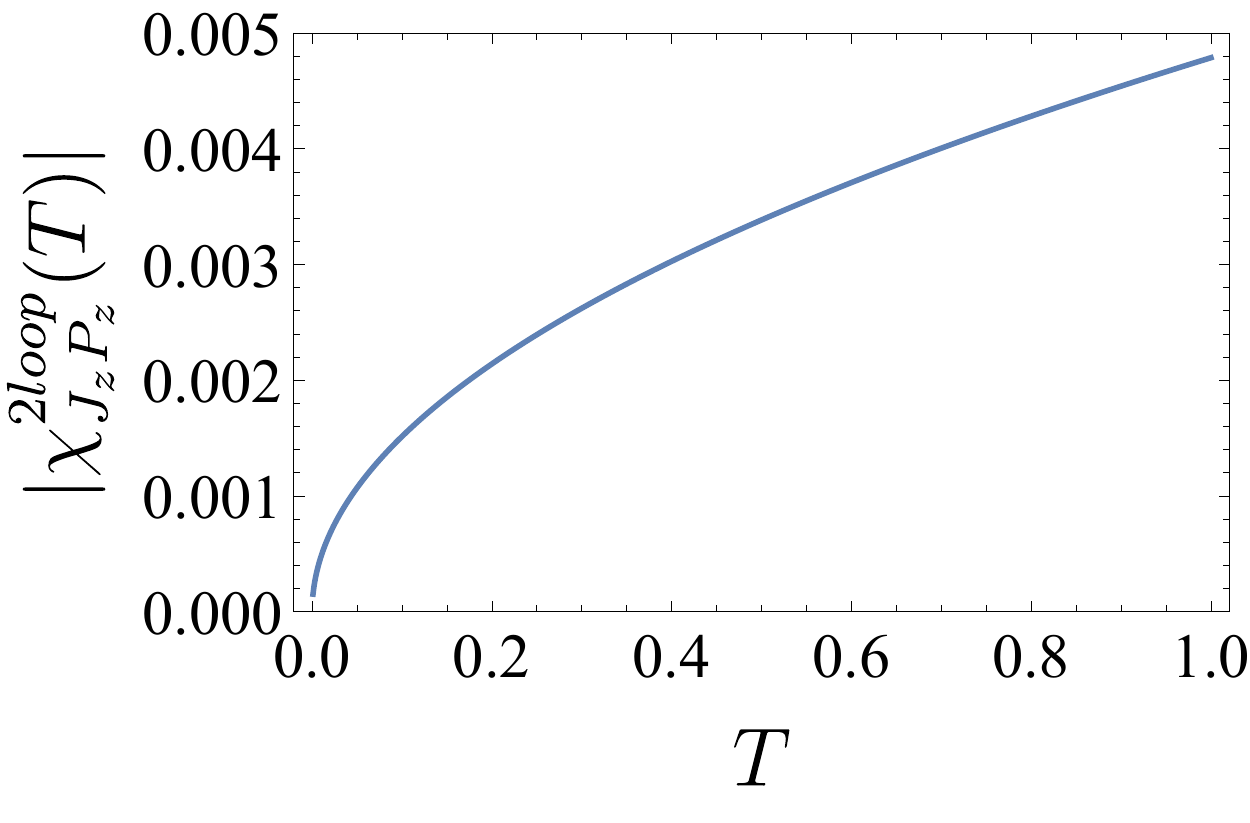}
\caption{\label{Fig:chi_JP_2loop}
Plot of the current-momentum susceptibility $\chi^{2loop}_{J_z P_z}(T)$ at two-loop order versus temperature $T$. Here, we have chosen the parameters $m=1$, $m'=5$, $e=0.1$, $c=1$, $N_f=1$, and $\Lambda_0=150$. The temperature dependence of this two-loop contribution is found to be $|\chi^{(2loop)}_{J_z P_z}(T)|\sim 0.005\,T^{1/2}$.}
\end{figure}

One can easily check that there are only three Feynman diagrams at two-loop order. The corresponding diagrams are similar to the ones in Figs.~\ref{fig2}, \ref{fig3}, and \ref{fig4}. These contributions evaluate to $\chi^{2loop}_{J_z P_z}(T)=\chi^{(2,1)}_{J_z P_z}(T)+\chi^{(2,2)}_{J_z P_z}(T)$, where
\begin{widetext}
\begin{align}
\label{chiJzPz}
&\chi^{(2,1)}_{J_z P_z}(T)\sim \left(\frac{8\,e^2 \Lambda_0}{15\pi^2 c\, T}\right)
T\,\sum \limits_{k_0}
\int \frac{d^3 {\mathbf k}}{(2\pi)^3}\, k_z \frac{\left[ \left (\mathrm{i}\,k_0-\frac{\mathbf{k}^2}{2m'}  \right )^3
\left( \partial_{k_z}\mathbf{d_k}\cdot \mathbf{d_k} \right)+3\left (\mathrm{i}\,k_0-\frac{\mathbf{k}^2}{2m'}  \right )
\left( \partial_{k_z}\mathbf{d_k}\cdot \mathbf{d_k}\right) |\mathbf{d_k}|^2 \right]}{\left[\left (\mathrm{i}\,k_0-\frac{\mathbf{k}^2}{2m'}  \right )^2-|\mathbf{d_k}|^2|\right]^3}\,,\\\label{chiJzPz-2}
&\chi^{(2,2)}_{J_z P_z}(T)\sim \left(\frac{e^2\,\Lambda_0}{2\pi^2 c\, T}\right)
T\,\sum \limits_{k_0}
\int \frac{d^3 {\mathbf k}}{(2\pi)^3}\, k_z\frac{\left (\mathrm{i}\,k_0-\frac{\mathbf{k}^2}{2m'}  \right )
\left( \partial_{k_z}\mathbf{d_k}\cdot \mathbf{d_k} \right)}
{\left[\left (\mathrm{i}\,k_0-\frac{\mathbf{k}^2}{2m'}  \right )^2-|\mathbf{d_k}|^2|\right]^2}
\,.
\end{align}
\end{widetext}
We provide the detailed steps of the calculation in Appendix~\ref{chi2loop}. Finally, we evaluate the expressions in Eqs.~\eqref{chiJzPz} and \eqref{chiJzPz-2} numerically, and obtain that $\chi^{2loop}_{J_zP_z}(T)\sim \frac{e^2}{c}\,T^{1/2}$ (see Fig. \ref{Fig:chi_JP_2loop}).

%%%%%%%%%%%%%%%%%%%%%%%%%%%%%%%%5
%%%%%%%%%%%%%%%%%5

%%%%%%%%%%%%%%%%%%%%%%%%%%%%%%%%%%%%
\subsection{Memory matrix calculation}

\begin{figure}[]
\includegraphics[width=0.17\textwidth]{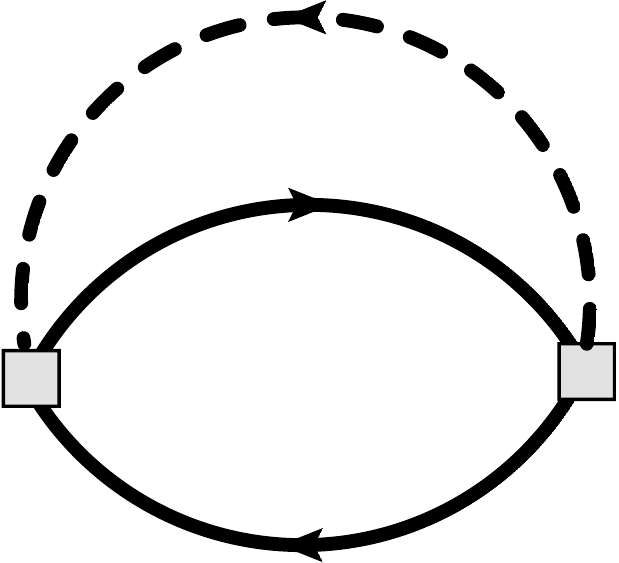}
\caption{\label{Fig:m_0}
Feynman diagram for the calculation of the leading-order contribution $M^{(0)}_{P_z P_z}(T)$ to the memory matrix. The solid line represents the bare fermionic propagator, whereas the dashed line represents the impurity line that carries only internal momentum and external energy $\omega$.} 
\end{figure}

\begin{figure}[]
\includegraphics[width=0.45\textwidth]{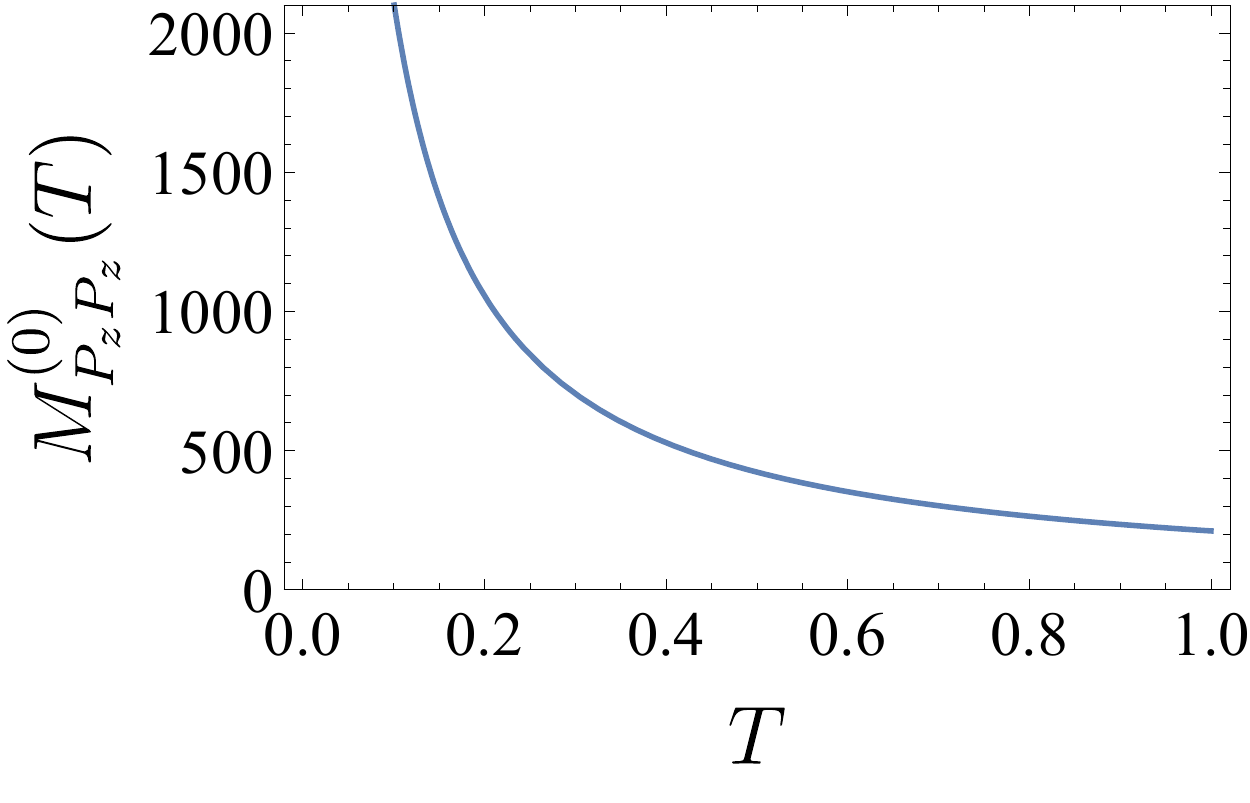}
\caption{\label{Fig:M_PP}
Plot of $M^{(0)}_{P_z P_z}(T)$ versus temperature $T$. Here, we have chosen the parameters $W_0=1$, $N_f=1$, $m=1$, $m'=5$, and $\Lambda_0=150$. To obtain the memory matrix, we have performed the analytical continuation $\mathrm{i}\,\omega\rightarrow \omega+\mathrm{i}\,\delta$, where we have set $\delta = 10^{-9}$. The curve corresponds to the fit given by $g(T)=a'+b'/T$, where the parameters $a'\approx  0.059$ and $b'\approx 211.25$ depend only on the UV cutoff $\Lambda_0$.} 
\end{figure}

We now compute the Feynman diagram associated with the calculation of the memory matrix to leading order, as shown in Fig.~\ref{Fig:m_0}, which is given by:
\begin{widetext}
\begin{align}
M^{(0)}_{P_z P_z}(T)&=-W_0 \, 
\lim_{\omega\rightarrow 0}\frac{\text{Im}
\bigg [
\int \frac{d^3 {\mathbf k} \, d^3 {\mathbf q}}{(2\pi)^6}
\,k_z^2\, T\sum \limits _{k_0}\text{Tr}\left[G_0(\omega+k_0,\mathbf{k+q})\,
G_0(k_0,\mathbf{k})\right]  \bigg ]
\Bigg|_{\mathrm{i} \,\omega\rightarrow\omega+\mathrm{i} \,\delta}}
{\omega}
\nonumber\\
&=-4\,W_0\,
\lim_{\omega\rightarrow 0}\frac{\text{Im}
\bigg[
\int \frac{d^3 {\mathbf k} \, d^3 {\mathbf q'}}{(2\pi)^6}\, k_z^2\, 
T\sum \limits _{k_0}
\frac{\left \lbrace \mathrm{i} \,k_0+\mathrm{i} \,\omega
-\frac{ \left(\mathbf{k+q} \right )^2}{2m'}\right \rbrace
\left(\mathrm{i} \,k_0-\frac{\mathbf k^2}{2m'}\right)
+ \left( \mathbf{d}_{\mathbf{k+q}}\cdot\mathbf{d}_{\mathbf{k}} \right)
}
{\left \lbrace \left(\mathrm{i} \,k_0
-\frac{(\mathbf{k+q})^2}{2m'}\right)^2-|\mathbf{d}_{\mathbf{k+q}}|^2\right \rbrace
\left \lbrace \left(\mathrm{i} \,k_0
-\frac{\mathbf k^2}{2m'}\right)^2-|\mathbf{d}_{\mathbf{k}}|^2
\right \rbrace }\bigg ]
\Bigg|_{\mathrm{i} \,\omega\rightarrow\omega+\mathrm{i} \,\delta}}{\omega}
\,.
\end{align}
\end{widetext}
As before, the summation over $k_0$ is evaluated using the method of residues. After performing the analytical continuation, the resulting integral is then evaluated numerically which finally gives $M^{(0)}_{P_z P_z}(T)/W_0\sim a'+b'/T$ (see Fig.~\ref{Fig:M_PP}), where $a'$ and $b'$ ($b'\gg a'$) are non-universal constants that depend only on the UV cutoff $\Lambda_0$. These constants are such that $a'$ scales as $\Lambda_0^2$ and $b'$ scales as $\Lambda_0^4$, leading to $\frac{a'} {b'}\rightarrow 0$ for $\Lambda_0\rightarrow \infty$. Therefore, the final expression can be effectively approximated as $M^{(0)}_{P_z P_z}(T)\approx b'/T$ at low temperatures.

%%%%%%%%%%%%%%%%%%%%%%
%%%%%%%%%%%%%%%%%%%%

%%%%%%%%%%%%%%%%%%%%%%%%%%%%%%
%%%%%%%%%%%%%%%%%%%%

%%%%%%%%%%%%%%%
\subsection{Scaling of dc conductivity}

Taking into account all contributions, the scaling of the dc conductivity of the LAB phase in the presence of weak short-ranged scalar disorder is given by:
\begin{align}
\sigma_{dc}(T)\equiv\frac{1}{\rho(T)}
=\frac{|\chi_{J_z P_z}(T)|^2}{M_{P_z P_z}(T)} \sim T^n\,, \text{ where }
2\lesssim n \lesssim 4\,,
\end{align}
and $\rho(T)$ is the resistivity.
It is important to compare this expression with the dc conductivity of the LAB phase in the clean limit.
If we assume that the $\omega/T$ scaling holds for the conductivity in this system,
then $\sigma_{dc}(T)\sim T^{\alpha^*}$ in the clean limit according to our optical conductivity results, where $\alpha^*\approx 0.54 $ is the renormalized exponent that violates hyperscaling for $d=3$ (i.e. $\varepsilon=1$) and $N_f=1$.
This implies that $\sigma_{dc}(T)$ in the presence of disorder displays a stronger power-law suppression as a function of temperature, which is an expected feature since the influence of disorder is a relevant perturbation in the vicinity of the LAB fixed point \cite{rahul-sid,ips-rahul,*ips-rahul-errata}. It is also interesting to compare our theoretical results with recent transport experiments \cite{Pramanik} performed on Luttinger semimetal compounds like pyrochlore iridates [(Y$_{1-x}$Pr$_x$)$_2$Ir$_2$O$_7$]. In these compounds, some degree of disorder is always present, and the dc resistivity has been found to follow the power-law $\rho(T)\sim T^{-n}$, with the exponent being $n\approx 2.98$ at zero doping  \cite{Pramanik}. Therefore, we conclude that our calculation is in qualitative agreement with these experimental data.

%%%%%%%%%%%%%%%%%%%%%%%%%%%%%%%%%%
%%%%%%%%%%%%%%%%%%%%%%%%%%%%%%%%%%%%%%%%

%%%%%%%%%%%%%%%%%%%%%%%
\section{Summary and outlook}
\label{end}

In this paper, we have computed the scaling behavior of the optical conductivity and the dc conductivity of the LAB phase of Luttinger semimetals, by means of the Kubo formula and the Mori-Zwanzig memory matrix method, respectively.
We have found that the optical conductivity in the LAB phase is characterized by a small violation (proportional to $\varepsilon=4-d$) of the hyperscaling property in the clean limit, in contrast to the low-energy effective theories that possess Dirac quasiparticles in the excitation spectrum (which obey hyperscaling).
In the computations for dc conductivity $\sigma_{dc}(T)$, we have included the effects of weak short-ranged scalar disorder. We have shown that $\sigma_{dc}(T)$ exhibits a stronger power-law suppression at low temperatures compared to the corresponding result in the clean limit. This was an expected feature since the influence of disorder is a relevant perturbation in the system. Lastly, we have directly compared our theoretical prediction with recent experiments performed in disordered Luttinger semimetal materials like the pyrochlore iridates \cite{Pramanik} and found qualitative agreement with the experimental data. In some other experiments \cite{armitage_expt}, the experimentalists have measured the optical conductivity in the Luttinger semimetal material Pr$_2$Ir$_2$O$_7$, but they could not tune the Fermi energy low enough to touch the band-crossing point. Their sample was thus a slightly doped Luttinger semimetal, where they found a number of signatures that are precursors to the LAB physics. Further experiments are planned in this direction, which will hopefully support our analytical findings. Moreover, from a theoretical point of view, it will be interesting to see if other computational strategies, such as the Kubo formula or the kinetic Boltzmann equation, are able to reproduce the dc conductivity at $T>0$ due to weak-disorder effects, which has been obtained here using the memory matrix approach.

We would like to point out that we have computed the finite-temperature scalings of the thermal conductivity and the thermoelectric coefficient of the LAB phase in a companion paper~\cite{ips-hermann2}. Finally, we would like to stress that it would be extremely interesting to investigate the effects of magnetic field on the magnetoresistance and the Hall coefficient of the LAB phase, and compare the results with the corresponding experimental data available for the pyrochlore iridates \cite{Pramanik}. The magnetic field breaks time-reversal symmetry and, in view of this, it must be a strongly relevant perturbation that ultimately makes the LAB fixed point unstable at low energy scales. We leave this analysis for future studies.

%.................................................
\begin{acknowledgments}

HF acknowledges funding from CNPq under Grant No. 310710/2018-9. 

%%%%%%%%%%%%
\end{acknowledgments}

%%%%%%%%%%%%%%%%%%%%%%%%%%%%
\bibliography{biblio}

\begin{thebibliography}{61}%
\makeatletter
\providecommand \@ifxundefined [1]{%
 \@ifx{#1\undefined}
}%
\providecommand \@ifnum [1]{%
 \ifnum #1\expandafter \@firstoftwo
 \else \expandafter \@secondoftwo
 \fi
}%
\providecommand \@ifx [1]{%
 \ifx #1\expandafter \@firstoftwo
 \else \expandafter \@secondoftwo
 \fi
}%
\providecommand \natexlab [1]{#1}%
\providecommand \enquote  [1]{``#1''}%
\providecommand \bibnamefont  [1]{#1}%
\providecommand \bibfnamefont [1]{#1}%
\providecommand \citenamefont [1]{#1}%
\providecommand \href@noop [0]{\@secondoftwo}%
\providecommand \href [0]{\begingroup \@sanitize@url \@href}%
\providecommand \@href[1]{\@@startlink{#1}\@@href}%
\providecommand \@@href[1]{\endgroup#1\@@endlink}%
\providecommand \@sanitize@url [0]{\catcode `\\12\catcode `\$12\catcode
  `\&12\catcode `\#12\catcode `\^12\catcode `\_12\catcode `\%12\relax}%
\providecommand \@@startlink[1]{}%
\providecommand \@@endlink[0]{}%
\providecommand \url  [0]{\begingroup\@sanitize@url \@url }%
\providecommand \@url [1]{\endgroup\@href {#1}{\urlprefix }}%
\providecommand \urlprefix  [0]{URL }%
\providecommand \Eprint [0]{\href }%
\providecommand \doibase [0]{https://doi.org/}%
\providecommand \selectlanguage [0]{\@gobble}%
\providecommand \bibinfo  [0]{\@secondoftwo}%
\providecommand \bibfield  [0]{\@secondoftwo}%
\providecommand \translation [1]{[#1]}%
\providecommand \BibitemOpen [0]{}%
\providecommand \bibitemStop [0]{}%
\providecommand \bibitemNoStop [0]{.\EOS\space}%
\providecommand \EOS [0]{\spacefactor3000\relax}%
\providecommand \BibitemShut  [1]{\csname bibitem#1\endcsname}%
\let\auto@bib@innerbib\@empty
%</preamble>
\bibitem [{\citenamefont {{Nayak}}\ and\ \citenamefont
  {{Wilczek}}(1994{\natexlab{a}})}]{nayak}%
  \BibitemOpen
  \bibfield  {author} {\bibinfo {author} {\bibfnamefont {C.}~\bibnamefont
  {{Nayak}}}\ and\ \bibinfo {author} {\bibfnamefont {F.}~\bibnamefont
  {{Wilczek}}},\ }\bibfield  {title} {\bibinfo {title} {{Renormalization group
  approach to low temperature properties of a non-Fermi liquid metal}},\ }\href
  {https://doi.org/10.1016/0550-3213(94)90158-9} {\bibfield  {journal}
  {\bibinfo  {journal} {Nuclear Physics B}\ }\textbf {\bibinfo {volume}
  {430}},\ \bibinfo {pages} {534} (\bibinfo {year}
  {1994}{\natexlab{a}})}\BibitemShut {NoStop}%
\bibitem [{\citenamefont {{Nayak}}\ and\ \citenamefont
  {{Wilczek}}(1994{\natexlab{b}})}]{nayak1}%
  \BibitemOpen
  \bibfield  {author} {\bibinfo {author} {\bibfnamefont {C.}~\bibnamefont
  {{Nayak}}}\ and\ \bibinfo {author} {\bibfnamefont {F.}~\bibnamefont
  {{Wilczek}}},\ }\bibfield  {title} {\bibinfo {title} {{Non-Fermi liquid fixed
  point in 2 + 1 dimensions}},\ }\href
  {https://doi.org/10.1016/0550-3213(94)90477-4} {\bibfield  {journal}
  {\bibinfo  {journal} {Nuclear Physics B}\ }\textbf {\bibinfo {volume}
  {417}},\ \bibinfo {pages} {359} (\bibinfo {year}
  {1994}{\natexlab{b}})}\BibitemShut {NoStop}%
\bibitem [{\citenamefont {{Lawler}}\ \emph {et~al.}(2006)\citenamefont
  {{Lawler}}, \citenamefont {{Barci}}, \citenamefont {{Fern{\'a}ndez}},
  \citenamefont {{Fradkin}},\ and\ \citenamefont {{Oxman}}}]{lawler1}%
  \BibitemOpen
  \bibfield  {author} {\bibinfo {author} {\bibfnamefont {M.~J.}\ \bibnamefont
  {{Lawler}}}, \bibinfo {author} {\bibfnamefont {D.~G.}\ \bibnamefont
  {{Barci}}}, \bibinfo {author} {\bibfnamefont {V.}~\bibnamefont
  {{Fern{\'a}ndez}}}, \bibinfo {author} {\bibfnamefont {E.}~\bibnamefont
  {{Fradkin}}},\ and\ \bibinfo {author} {\bibfnamefont {L.}~\bibnamefont
  {{Oxman}}},\ }\bibfield  {title} {\bibinfo {title} {{Nonperturbative behavior
  of the quantum phase transition to a nematic Fermi fluid}},\ }\href
  {https://doi.org/10.1103/PhysRevB.73.085101} {\bibfield  {journal} {\bibinfo
  {journal} {\prb}\ }\textbf {\bibinfo {volume} {73}},\ \bibinfo {eid} {085101}
  (\bibinfo {year} {2006})}\BibitemShut {NoStop}%
\bibitem [{\citenamefont {Mross}\ \emph {et~al.}(2010)\citenamefont {Mross},
  \citenamefont {McGreevy}, \citenamefont {Liu},\ and\ \citenamefont
  {Senthil}}]{mross}%
  \BibitemOpen
  \bibfield  {author} {\bibinfo {author} {\bibfnamefont {D.~F.}\ \bibnamefont
  {Mross}}, \bibinfo {author} {\bibfnamefont {J.}~\bibnamefont {McGreevy}},
  \bibinfo {author} {\bibfnamefont {H.}~\bibnamefont {Liu}},\ and\ \bibinfo
  {author} {\bibfnamefont {T.}~\bibnamefont {Senthil}},\ }\bibfield  {title}
  {\bibinfo {title} {Controlled expansion for certain non-fermi-liquid
  metals},\ }\href {https://doi.org/10.1103/PhysRevB.82.045121} {\bibfield
  {journal} {\bibinfo  {journal} {Phys. Rev. B}\ }\textbf {\bibinfo {volume}
  {82}},\ \bibinfo {pages} {045121} (\bibinfo {year} {2010})}\BibitemShut
  {NoStop}%
\bibitem [{\citenamefont {{Jiang}}\ \emph {et~al.}(2013)\citenamefont
  {{Jiang}}, \citenamefont {{Block}}, \citenamefont {{Mishmash}}, \citenamefont
  {{Garrison}}, \citenamefont {{Sheng}}, \citenamefont {{Motrunich}},\ and\
  \citenamefont {{Fisher}}}]{Jiang}%
  \BibitemOpen
  \bibfield  {author} {\bibinfo {author} {\bibfnamefont {H.-C.}\ \bibnamefont
  {{Jiang}}}, \bibinfo {author} {\bibfnamefont {M.~S.}\ \bibnamefont
  {{Block}}}, \bibinfo {author} {\bibfnamefont {R.~V.}\ \bibnamefont
  {{Mishmash}}}, \bibinfo {author} {\bibfnamefont {J.~R.}\ \bibnamefont
  {{Garrison}}}, \bibinfo {author} {\bibfnamefont {D.~N.}\ \bibnamefont
  {{Sheng}}}, \bibinfo {author} {\bibfnamefont {O.~I.}\ \bibnamefont
  {{Motrunich}}},\ and\ \bibinfo {author} {\bibfnamefont {M.~P.~A.}\
  \bibnamefont {{Fisher}}},\ }\bibfield  {title} {\bibinfo {title}
  {{Non-Fermi-liquid d-wave metal phase of strongly interacting electrons}},\
  }\href {https://doi.org/10.1038/nature11732} {\bibfield  {journal} {\bibinfo
  {journal} {\nat}\ }\textbf {\bibinfo {volume} {493}},\ \bibinfo {pages} {39}
  (\bibinfo {year} {2013})}\BibitemShut {NoStop}%
\bibitem [{\citenamefont {Chung}\ \emph {et~al.}(2013)\citenamefont {Chung},
  \citenamefont {Mandal}, \citenamefont {Raghu},\ and\ \citenamefont
  {Chakravarty}}]{ips2}%
  \BibitemOpen
  \bibfield  {author} {\bibinfo {author} {\bibfnamefont {S.~B.}\ \bibnamefont
  {Chung}}, \bibinfo {author} {\bibfnamefont {I.}~\bibnamefont {Mandal}},
  \bibinfo {author} {\bibfnamefont {S.}~\bibnamefont {Raghu}},\ and\ \bibinfo
  {author} {\bibfnamefont {S.}~\bibnamefont {Chakravarty}},\ }\bibfield
  {title} {\bibinfo {title} {Higher angular momentum pairing from transverse
  gauge interactions},\ }\href {https://doi.org/10.1103/PhysRevB.88.045127}
  {\bibfield  {journal} {\bibinfo  {journal} {Phys. Rev. B}\ }\textbf {\bibinfo
  {volume} {88}},\ \bibinfo {pages} {045127} (\bibinfo {year}
  {2013})}\BibitemShut {NoStop}%
\bibitem [{\citenamefont {Wang}\ \emph {et~al.}(2014)\citenamefont {Wang},
  \citenamefont {Mandal}, \citenamefont {Chung},\ and\ \citenamefont
  {Chakravarty}}]{ips3}%
  \BibitemOpen
  \bibfield  {author} {\bibinfo {author} {\bibfnamefont {Z.}~\bibnamefont
  {Wang}}, \bibinfo {author} {\bibfnamefont {I.}~\bibnamefont {Mandal}},
  \bibinfo {author} {\bibfnamefont {S.~B.}\ \bibnamefont {Chung}},\ and\
  \bibinfo {author} {\bibfnamefont {S.}~\bibnamefont {Chakravarty}},\
  }\bibfield  {title} {\bibinfo {title} {Pairing in half-filled landau level},\
  }\href {https://doi.org/http://dx.doi.org/10.1016/j.aop.2014.09.021}
  {\bibfield  {journal} {\bibinfo  {journal} {Annals of Physics}\ }\textbf
  {\bibinfo {volume} {351}},\ \bibinfo {pages} {727 } (\bibinfo {year}
  {2014})}\BibitemShut {NoStop}%
\bibitem [{\citenamefont {Sur}\ and\ \citenamefont {Lee}(2014)}]{Shouvik1}%
  \BibitemOpen
  \bibfield  {author} {\bibinfo {author} {\bibfnamefont {S.}~\bibnamefont
  {Sur}}\ and\ \bibinfo {author} {\bibfnamefont {S.-S.}\ \bibnamefont {Lee}},\
  }\bibfield  {title} {\bibinfo {title} {Chiral non-fermi liquids},\ }\href
  {https://doi.org/10.1103/PhysRevB.90.045121} {\bibfield  {journal} {\bibinfo
  {journal} {Phys. Rev. B}\ }\textbf {\bibinfo {volume} {90}},\ \bibinfo
  {pages} {045121} (\bibinfo {year} {2014})}\BibitemShut {NoStop}%
\bibitem [{\citenamefont {Dalidovich}\ and\ \citenamefont
  {Lee}(2013)}]{Lee-Dalid}%
  \BibitemOpen
  \bibfield  {author} {\bibinfo {author} {\bibfnamefont {D.}~\bibnamefont
  {Dalidovich}}\ and\ \bibinfo {author} {\bibfnamefont {S.-S.}\ \bibnamefont
  {Lee}},\ }\bibfield  {title} {\bibinfo {title} {Perturbative non-fermi
  liquids from dimensional regularization},\ }\href
  {https://doi.org/10.1103/PhysRevB.88.245106} {\bibfield  {journal} {\bibinfo
  {journal} {Phys. Rev. B}\ }\textbf {\bibinfo {volume} {88}},\ \bibinfo
  {pages} {245106} (\bibinfo {year} {2013})}\BibitemShut {NoStop}%
\bibitem [{\citenamefont {Sur}\ and\ \citenamefont {Lee}(2015)}]{shouvik2}%
  \BibitemOpen
  \bibfield  {author} {\bibinfo {author} {\bibfnamefont {S.}~\bibnamefont
  {Sur}}\ and\ \bibinfo {author} {\bibfnamefont {S.-S.}\ \bibnamefont {Lee}},\
  }\bibfield  {title} {\bibinfo {title} {Quasilocal strange metal},\ }\href
  {https://doi.org/10.1103/PhysRevB.91.125136} {\bibfield  {journal} {\bibinfo
  {journal} {Phys. Rev. B}\ }\textbf {\bibinfo {volume} {91}},\ \bibinfo
  {pages} {125136} (\bibinfo {year} {2015})}\BibitemShut {NoStop}%
\bibitem [{\citenamefont {de~Carvalho}\ \emph {et~al.}(2015)\citenamefont
  {de~Carvalho}, \citenamefont {Kloss}, \citenamefont {Montiel}, \citenamefont
  {Freire},\ and\ \citenamefont {P\'epin}}]{Freire_Pepin_2}%
  \BibitemOpen
  \bibfield  {author} {\bibinfo {author} {\bibfnamefont {V.~S.}\ \bibnamefont
  {de~Carvalho}}, \bibinfo {author} {\bibfnamefont {T.}~\bibnamefont {Kloss}},
  \bibinfo {author} {\bibfnamefont {X.}~\bibnamefont {Montiel}}, \bibinfo
  {author} {\bibfnamefont {H.}~\bibnamefont {Freire}},\ and\ \bibinfo {author}
  {\bibfnamefont {C.}~\bibnamefont {P\'epin}},\ }\bibfield  {title} {\bibinfo
  {title} {Strong competition between
  ${\mathrm{\ensuremath{\Theta}}}_{II}$-loop-current order and $d$-wave charge
  order along the diagonal direction in a two-dimensional hot spot model},\
  }\href {https://link.aps.org/doi/10.1103/PhysRevB.92.075123} {\bibfield
  {journal} {\bibinfo  {journal} {Phys. Rev. B}\ }\textbf {\bibinfo {volume}
  {92}},\ \bibinfo {pages} {075123} (\bibinfo {year} {2015})}\BibitemShut
  {NoStop}%
\bibitem [{\citenamefont {Mandal}\ and\ \citenamefont
  {Lee}(2015)}]{ips-uv-ir1}%
  \BibitemOpen
  \bibfield  {author} {\bibinfo {author} {\bibfnamefont {I.}~\bibnamefont
  {Mandal}}\ and\ \bibinfo {author} {\bibfnamefont {S.-S.}\ \bibnamefont
  {Lee}},\ }\bibfield  {title} {\bibinfo {title} {Ultraviolet/infrared mixing
  in non-fermi liquids},\ }\href {https://doi.org/10.1103/PhysRevB.92.035141}
  {\bibfield  {journal} {\bibinfo  {journal} {Phys. Rev. B}\ }\textbf {\bibinfo
  {volume} {92}},\ \bibinfo {pages} {035141} (\bibinfo {year}
  {2015})}\BibitemShut {NoStop}%
\bibitem [{\citenamefont {de~Carvalho}\ \emph {et~al.}(2016)\citenamefont
  {de~Carvalho}, \citenamefont {P\'epin},\ and\ \citenamefont
  {Freire}}]{Freire_Pepin_1}%
  \BibitemOpen
  \bibfield  {author} {\bibinfo {author} {\bibfnamefont {V.~S.}\ \bibnamefont
  {de~Carvalho}}, \bibinfo {author} {\bibfnamefont {C.}~\bibnamefont
  {P\'epin}},\ and\ \bibinfo {author} {\bibfnamefont {H.}~\bibnamefont
  {Freire}},\ }\bibfield  {title} {\bibinfo {title} {Coexistence of
  ${\mathrm{\ensuremath{\Theta}}}_{II}$-loop-current order with checkerboard
  $d$-wave cdw/pdw order in a hot-spot model for cuprate superconductors},\
  }\href {https://link.aps.org/doi/10.1103/PhysRevB.93.115144} {\bibfield
  {journal} {\bibinfo  {journal} {Phys. Rev. B}\ }\textbf {\bibinfo {volume}
  {93}},\ \bibinfo {pages} {115144} (\bibinfo {year} {2016})}\BibitemShut
  {NoStop}%
\bibitem [{\citenamefont {Mandal}(2016{\natexlab{a}})}]{ips-uv-ir2}%
  \BibitemOpen
  \bibfield  {author} {\bibinfo {author} {\bibfnamefont {I.}~\bibnamefont
  {Mandal}},\ }\bibfield  {title} {\bibinfo {title} {{UV/IR Mixing In Non-Fermi
  Liquids: Higher-Loop Corrections In Different Energy Ranges}},\ }\href
  {https://doi.org/10.1140/epjb/e2016-70509-4} {\bibfield  {journal} {\bibinfo
  {journal} {Eur. Phys. J. B}\ }\textbf {\bibinfo {volume} {89}},\ \bibinfo
  {pages} {278} (\bibinfo {year} {2016}{\natexlab{a}})}\BibitemShut {NoStop}%
\bibitem [{\citenamefont {Eberlein}\ \emph {et~al.}(2016)\citenamefont
  {Eberlein}, \citenamefont {Mandal},\ and\ \citenamefont
  {Sachdev}}]{ips-subir}%
  \BibitemOpen
  \bibfield  {author} {\bibinfo {author} {\bibfnamefont {A.}~\bibnamefont
  {Eberlein}}, \bibinfo {author} {\bibfnamefont {I.}~\bibnamefont {Mandal}},\
  and\ \bibinfo {author} {\bibfnamefont {S.}~\bibnamefont {Sachdev}},\
  }\bibfield  {title} {\bibinfo {title} {Hyperscaling violation at the
  ising-nematic quantum critical point in two-dimensional metals},\ }\href
  {https://doi.org/10.1103/PhysRevB.94.045133} {\bibfield  {journal} {\bibinfo
  {journal} {Phys. Rev. B}\ }\textbf {\bibinfo {volume} {94}},\ \bibinfo
  {pages} {045133} (\bibinfo {year} {2016})}\BibitemShut {NoStop}%
\bibitem [{\citenamefont {Mandal}(2016{\natexlab{b}})}]{ips-sc}%
  \BibitemOpen
  \bibfield  {author} {\bibinfo {author} {\bibfnamefont {I.}~\bibnamefont
  {Mandal}},\ }\bibfield  {title} {\bibinfo {title} {Superconducting
  instability in non-fermi liquids},\ }\href
  {https://doi.org/10.1103/PhysRevB.94.115138} {\bibfield  {journal} {\bibinfo
  {journal} {Phys. Rev. B}\ }\textbf {\bibinfo {volume} {94}},\ \bibinfo
  {pages} {115138} (\bibinfo {year} {2016}{\natexlab{b}})}\BibitemShut
  {NoStop}%
\bibitem [{\citenamefont {Mandal}(2017)}]{ips-c2}%
  \BibitemOpen
  \bibfield  {author} {\bibinfo {author} {\bibfnamefont {I.}~\bibnamefont
  {Mandal}},\ }\bibfield  {title} {\bibinfo {title} {Scaling behaviour and
  superconducting instability in anisotropic non-fermi liquids},\ }\href
  {https://doi.org/https://doi.org/10.1016/j.aop.2016.11.009} {\bibfield
  {journal} {\bibinfo  {journal} {Annals of Physics}\ }\textbf {\bibinfo
  {volume} {376}},\ \bibinfo {pages} {89 } (\bibinfo {year}
  {2017})}\BibitemShut {NoStop}%
\bibitem [{\citenamefont {Lee}(2018)}]{Lee_2018}%
  \BibitemOpen
  \bibfield  {author} {\bibinfo {author} {\bibfnamefont {S.-S.}\ \bibnamefont
  {Lee}},\ }\bibfield  {title} {\bibinfo {title} {Recent developments in
  non-fermi liquid theory},\ }\href
  {https://doi.org/10.1146/annurev-conmatphys-031016-025531} {\bibfield
  {journal} {\bibinfo  {journal} {Annual Review of Condensed Matter Physics}\
  }\textbf {\bibinfo {volume} {9}},\ \bibinfo {pages} {227–244} (\bibinfo
  {year} {2018})}\BibitemShut {NoStop}%
\bibitem [{\citenamefont {Pimenov}\ \emph {et~al.}(2018)\citenamefont
  {Pimenov}, \citenamefont {Mandal}, \citenamefont {Piazza},\ and\
  \citenamefont {Punk}}]{ips-fflo}%
  \BibitemOpen
  \bibfield  {author} {\bibinfo {author} {\bibfnamefont {D.}~\bibnamefont
  {Pimenov}}, \bibinfo {author} {\bibfnamefont {I.}~\bibnamefont {Mandal}},
  \bibinfo {author} {\bibfnamefont {F.}~\bibnamefont {Piazza}},\ and\ \bibinfo
  {author} {\bibfnamefont {M.}~\bibnamefont {Punk}},\ }\bibfield  {title}
  {\bibinfo {title} {Non-fermi liquid at the fflo quantum critical point},\
  }\href {https://doi.org/10.1103/PhysRevB.98.024510} {\bibfield  {journal}
  {\bibinfo  {journal} {Phys. Rev. B}\ }\textbf {\bibinfo {volume} {98}},\
  \bibinfo {pages} {024510} (\bibinfo {year} {2018})}\BibitemShut {NoStop}%
\bibitem [{\citenamefont {Mandal}(2020{\natexlab{a}})}]{ips-nfl-u1}%
  \BibitemOpen
  \bibfield  {author} {\bibinfo {author} {\bibfnamefont {I.}~\bibnamefont
  {Mandal}},\ }\bibfield  {title} {\bibinfo {title} {Critical fermi surfaces in
  generic dimensions arising from transverse gauge field interactions},\ }\href
  {https://doi.org/10.1103/PhysRevResearch.2.043277} {\bibfield  {journal}
  {\bibinfo  {journal} {Phys. Rev. Research}\ }\textbf {\bibinfo {volume}
  {2}},\ \bibinfo {pages} {043277} (\bibinfo {year}
  {2020}{\natexlab{a}})}\BibitemShut {NoStop}%
\bibitem [{\citenamefont {Moon}\ \emph {et~al.}(2013)\citenamefont {Moon},
  \citenamefont {Xu}, \citenamefont {Kim},\ and\ \citenamefont
  {Balents}}]{moon-xu}%
  \BibitemOpen
  \bibfield  {author} {\bibinfo {author} {\bibfnamefont {E.-G.}\ \bibnamefont
  {Moon}}, \bibinfo {author} {\bibfnamefont {C.}~\bibnamefont {Xu}}, \bibinfo
  {author} {\bibfnamefont {Y.~B.}\ \bibnamefont {Kim}},\ and\ \bibinfo {author}
  {\bibfnamefont {L.}~\bibnamefont {Balents}},\ }\bibfield  {title} {\bibinfo
  {title} {Non-fermi-liquid and topological states with strong spin-orbit
  coupling},\ }\href {https://doi.org/10.1103/PhysRevLett.111.206401}
  {\bibfield  {journal} {\bibinfo  {journal} {Phys. Rev. Lett.}\ }\textbf
  {\bibinfo {volume} {111}},\ \bibinfo {pages} {206401} (\bibinfo {year}
  {2013})}\BibitemShut {NoStop}%
\bibitem [{\citenamefont {Nandkishore}\ and\ \citenamefont
  {Parameswaran}(2017)}]{rahul-sid}%
  \BibitemOpen
  \bibfield  {author} {\bibinfo {author} {\bibfnamefont {R.~M.}\ \bibnamefont
  {Nandkishore}}\ and\ \bibinfo {author} {\bibfnamefont {S.~A.}\ \bibnamefont
  {Parameswaran}},\ }\bibfield  {title} {\bibinfo {title} {Disorder-driven
  destruction of a non-fermi liquid semimetal studied by renormalization group
  analysis},\ }\href {https://doi.org/10.1103/PhysRevB.95.205106} {\bibfield
  {journal} {\bibinfo  {journal} {Phys. Rev. B}\ }\textbf {\bibinfo {volume}
  {95}},\ \bibinfo {pages} {205106} (\bibinfo {year} {2017})}\BibitemShut
  {NoStop}%
\bibitem [{\citenamefont {{Mandal}}\ and\ \citenamefont
  {{Nandkishore}}(2018)}]{ips-rahul}%
  \BibitemOpen
  \bibfield  {author} {\bibinfo {author} {\bibfnamefont {I.}~\bibnamefont
  {{Mandal}}}\ and\ \bibinfo {author} {\bibfnamefont {R.~M.}\ \bibnamefont
  {{Nandkishore}}},\ }\bibfield  {title} {\bibinfo {title} {{Interplay of
  Coulomb interactions and disorder in three-dimensional quadratic band
  crossings without time-reversal symmetry and with unequal masses for
  conduction and valence bands}},\ }\href
  {https://doi.org/10.1103/PhysRevB.97.125121} {\bibfield  {journal} {\bibinfo
  {journal} {\prb}\ }\textbf {\bibinfo {volume} {97}},\ \bibinfo {eid} {125121}
  (\bibinfo {year} {2018})}\BibitemShut {NoStop}%
\bibitem [{\citenamefont {Mandal}\ and\ \citenamefont
  {Nandkishore}(2022)}]{ips-rahul-errata}%
  \BibitemOpen
  \bibfield  {author} {\bibinfo {author} {\bibfnamefont {I.}~\bibnamefont
  {Mandal}}\ and\ \bibinfo {author} {\bibfnamefont {R.~M.}\ \bibnamefont
  {Nandkishore}},\ }\bibfield  {title} {\bibinfo {title} {Erratum: Interplay of
  coulomb interactions and disorder in three-dimensional quadratic band
  crossings without time-reversal symmetry and with unequal masses for
  conduction and valence bands [{P}hys. {R}ev. {B} 97, 125121 (2018)]},\ }\href
  {https://doi.org/10.1103/PhysRevB.105.039901} {\bibfield  {journal} {\bibinfo
   {journal} {Phys. Rev. B}\ }\textbf {\bibinfo {volume} {105}},\ \bibinfo
  {pages} {039901} (\bibinfo {year} {2022})}\BibitemShut {NoStop}%
\bibitem [{\citenamefont {Mandal}(2018)}]{ips-qbt-sc}%
  \BibitemOpen
  \bibfield  {author} {\bibinfo {author} {\bibfnamefont {I.}~\bibnamefont
  {Mandal}},\ }\bibfield  {title} {\bibinfo {title} {Fate of superconductivity
  in three-dimensional disordered luttinger semimetals},\ }\href
  {https://doi.org/https://doi.org/10.1016/j.aop.2018.03.004} {\bibfield
  {journal} {\bibinfo  {journal} {Annals of Physics}\ }\textbf {\bibinfo
  {volume} {392}},\ \bibinfo {pages} {179 } (\bibinfo {year}
  {2018})}\BibitemShut {NoStop}%
\bibitem [{\citenamefont {Mandal}(2019)}]{ips_qbt_plasmons}%
  \BibitemOpen
  \bibfield  {author} {\bibinfo {author} {\bibfnamefont {I.}~\bibnamefont
  {Mandal}},\ }\bibfield  {title} {\bibinfo {title} {Search for plasmons in
  isotropic luttinger semimetals},\ }\href
  {https://doi.org/10.1016/j.aop.2019.04.002} {\bibfield  {journal} {\bibinfo
  {journal} {Annals of Physics}\ }\textbf {\bibinfo {volume} {406}},\ \bibinfo
  {pages} {173–185} (\bibinfo {year} {2019})}\BibitemShut {NoStop}%
\bibitem [{\citenamefont {Mandal}(2020{\natexlab{b}})}]{ips_qbt_tunnel}%
  \BibitemOpen
  \bibfield  {author} {\bibinfo {author} {\bibfnamefont {I.}~\bibnamefont
  {Mandal}},\ }\bibfield  {title} {\bibinfo {title} {Tunneling in fermi systems
  with quadratic band crossing points},\ }\href
  {https://doi.org/10.1016/j.aop.2020.168235} {\bibfield  {journal} {\bibinfo
  {journal} {Annals of Physics}\ }\textbf {\bibinfo {volume} {419}},\ \bibinfo
  {pages} {168235} (\bibinfo {year} {2020}{\natexlab{b}})}\BibitemShut
  {NoStop}%
\bibitem [{\citenamefont {Abrikosov}(1974)}]{Abrikosov}%
  \BibitemOpen
  \bibfield  {author} {\bibinfo {author} {\bibfnamefont {A.~A.}\ \bibnamefont
  {Abrikosov}},\ }\bibfield  {title} {\bibinfo {title} {Calculation of critical
  indices for zero-gap semiconductors},\ }\href@noop {} {\bibfield  {journal}
  {\bibinfo  {journal} {Sov. Phys.-JETP}\ }\textbf {\bibinfo {volume} {39}},\
  \bibinfo {pages} {709} (\bibinfo {year} {1974})}\BibitemShut {NoStop}%
\bibitem [{\citenamefont {Link}\ and\ \citenamefont
  {Herbut}(2020)}]{Herbut-PRB}%
  \BibitemOpen
  \bibfield  {author} {\bibinfo {author} {\bibfnamefont {J.~M.}\ \bibnamefont
  {Link}}\ and\ \bibinfo {author} {\bibfnamefont {I.~F.}\ \bibnamefont
  {Herbut}},\ }\bibfield  {title} {\bibinfo {title} {Hydrodynamic transport in
  the luttinger-abrikosov-beneslavskii non-fermi liquid},\ }\href
  {https://doi.org/10.1103/PhysRevB.101.125128} {\bibfield  {journal} {\bibinfo
   {journal} {Phys. Rev. B}\ }\textbf {\bibinfo {volume} {101}},\ \bibinfo
  {pages} {125128} (\bibinfo {year} {2020})}\BibitemShut {NoStop}%
\bibitem [{\citenamefont {Kovtun}\ \emph {et~al.}(2005)\citenamefont {Kovtun},
  \citenamefont {Son},\ and\ \citenamefont {Starinets}}]{DTSon-PRL_2005}%
  \BibitemOpen
  \bibfield  {author} {\bibinfo {author} {\bibfnamefont {P.~K.}\ \bibnamefont
  {Kovtun}}, \bibinfo {author} {\bibfnamefont {D.~T.}\ \bibnamefont {Son}},\
  and\ \bibinfo {author} {\bibfnamefont {A.~O.}\ \bibnamefont {Starinets}},\
  }\bibfield  {title} {\bibinfo {title} {Viscosity in strongly interacting
  quantum field theories from black hole physics},\ }\href
  {https://doi.org/10.1103/PhysRevLett.94.111601} {\bibfield  {journal}
  {\bibinfo  {journal} {Phys. Rev. Lett.}\ }\textbf {\bibinfo {volume} {94}},\
  \bibinfo {pages} {111601} (\bibinfo {year} {2005})}\BibitemShut {NoStop}%
\bibitem [{\citenamefont {Fritz}\ \emph {et~al.}(2008)\citenamefont {Fritz},
  \citenamefont {Schmalian}, \citenamefont {M\"uller},\ and\ \citenamefont
  {Sachdev}}]{Fritz-PRB}%
  \BibitemOpen
  \bibfield  {author} {\bibinfo {author} {\bibfnamefont {L.}~\bibnamefont
  {Fritz}}, \bibinfo {author} {\bibfnamefont {J.}~\bibnamefont {Schmalian}},
  \bibinfo {author} {\bibfnamefont {M.}~\bibnamefont {M\"uller}},\ and\
  \bibinfo {author} {\bibfnamefont {S.}~\bibnamefont {Sachdev}},\ }\bibfield
  {title} {\bibinfo {title} {Quantum critical transport in clean graphene},\
  }\href {https://doi.org/10.1103/PhysRevB.78.085416} {\bibfield  {journal}
  {\bibinfo  {journal} {Phys. Rev. B}\ }\textbf {\bibinfo {volume} {78}},\
  \bibinfo {pages} {085416} (\bibinfo {year} {2008})}\BibitemShut {NoStop}%
\bibitem [{\citenamefont {Policastro}\ \emph {et~al.}(2001)\citenamefont
  {Policastro}, \citenamefont {Son},\ and\ \citenamefont
  {Starinets}}]{DTSon-PRL}%
  \BibitemOpen
  \bibfield  {author} {\bibinfo {author} {\bibfnamefont {G.}~\bibnamefont
  {Policastro}}, \bibinfo {author} {\bibfnamefont {D.~T.}\ \bibnamefont
  {Son}},\ and\ \bibinfo {author} {\bibfnamefont {A.~O.}\ \bibnamefont
  {Starinets}},\ }\bibfield  {title} {\bibinfo {title} {Shear viscosity of
  strongly coupled n=4 supersymmetric yang-mills plasma},\ }\href
  {https://doi.org/10.1103/PhysRevLett.87.081601} {\bibfield  {journal}
  {\bibinfo  {journal} {Phys. Rev. Lett.}\ }\textbf {\bibinfo {volume} {87}},\
  \bibinfo {pages} {081601} (\bibinfo {year} {2001})}\BibitemShut {NoStop}%
\bibitem [{\citenamefont {Cao}\ \emph {et~al.}(2011)\citenamefont {Cao},
  \citenamefont {Elliott}, \citenamefont {Joseph}, \citenamefont {Wu},
  \citenamefont {Petricka}, \citenamefont {Sch{\"a}fer},\ and\ \citenamefont
  {Thomas}}]{Cao}%
  \BibitemOpen
  \bibfield  {author} {\bibinfo {author} {\bibfnamefont {C.}~\bibnamefont
  {Cao}}, \bibinfo {author} {\bibfnamefont {E.}~\bibnamefont {Elliott}},
  \bibinfo {author} {\bibfnamefont {J.}~\bibnamefont {Joseph}}, \bibinfo
  {author} {\bibfnamefont {H.}~\bibnamefont {Wu}}, \bibinfo {author}
  {\bibfnamefont {J.}~\bibnamefont {Petricka}}, \bibinfo {author}
  {\bibfnamefont {T.}~\bibnamefont {Sch{\"a}fer}},\ and\ \bibinfo {author}
  {\bibfnamefont {J.~E.}\ \bibnamefont {Thomas}},\ }\bibfield  {title}
  {\bibinfo {title} {Universal quantum viscosity in a unitary fermi gas},\
  }\href {https://doi.org/10.1126/science.1195219} {\bibfield  {journal}
  {\bibinfo  {journal} {Science}\ }\textbf {\bibinfo {volume} {331}},\ \bibinfo
  {pages} {58} (\bibinfo {year} {2011})}\BibitemShut {NoStop}%
\bibitem [{\citenamefont {Luttinger}(1956)}]{Luttinger}%
  \BibitemOpen
  \bibfield  {author} {\bibinfo {author} {\bibfnamefont {J.~M.}\ \bibnamefont
  {Luttinger}},\ }\bibfield  {title} {\bibinfo {title} {Quantum theory of
  cyclotron resonance in semiconductors: General theory},\ }\href
  {https://link.aps.org/doi/10.1103/PhysRev.102.1030} {\bibfield  {journal}
  {\bibinfo  {journal} {Phys. Rev.}\ }\textbf {\bibinfo {volume} {102}},\
  \bibinfo {pages} {1030} (\bibinfo {year} {1956})}\BibitemShut {NoStop}%
\bibitem [{\citenamefont {Murakami}\ \emph {et~al.}(2004)\citenamefont
  {Murakami}, \citenamefont {Nagosa},\ and\ \citenamefont {Zhang}}]{murakami}%
  \BibitemOpen
  \bibfield  {author} {\bibinfo {author} {\bibfnamefont {S.}~\bibnamefont
  {Murakami}}, \bibinfo {author} {\bibfnamefont {N.}~\bibnamefont {Nagosa}},\
  and\ \bibinfo {author} {\bibfnamefont {S.-C.}\ \bibnamefont {Zhang}},\
  }\bibfield  {title} {\bibinfo {title} {$\text{SU}(2)$ non-abelian holonomy
  and dissipationless spin current in semiconductors},\ }\href
  {https://link.aps.org/doi/10.1103/PhysRevB.69.235206} {\bibfield  {journal}
  {\bibinfo  {journal} {Phys. Rev. B}\ }\textbf {\bibinfo {volume} {69}},\
  \bibinfo {pages} {235206} (\bibinfo {year} {2004})}\BibitemShut {NoStop}%
\bibitem [{\citenamefont {Boettcher}\ and\ \citenamefont
  {Herbut}(2016)}]{igor16}%
  \BibitemOpen
  \bibfield  {author} {\bibinfo {author} {\bibfnamefont {I.}~\bibnamefont
  {Boettcher}}\ and\ \bibinfo {author} {\bibfnamefont {I.~F.}\ \bibnamefont
  {Herbut}},\ }\bibfield  {title} {\bibinfo {title} {Superconducting quantum
  criticality in three-dimensional luttinger semimetals},\ }\href
  {https://doi.org/10.1103/PhysRevB.93.205138} {\bibfield  {journal} {\bibinfo
  {journal} {Phys. Rev. B}\ }\textbf {\bibinfo {volume} {93}},\ \bibinfo
  {pages} {205138} (\bibinfo {year} {2016})}\BibitemShut {NoStop}%
\bibitem [{\citenamefont {Peskin}\ and\ \citenamefont
  {Schroeder}(1995)}]{Peskin}%
  \BibitemOpen
  \bibfield  {author} {\bibinfo {author} {\bibfnamefont {M.~E.}\ \bibnamefont
  {Peskin}}\ and\ \bibinfo {author} {\bibfnamefont {D.~V.}\ \bibnamefont
  {Schroeder}},\ }\href@noop {} {\emph {\bibinfo {title} {{An Introduction to
  Quantum Field Theory}}}}\ (\bibinfo  {publisher} {Addison-Wesley},\ \bibinfo
  {address} {Reading},\ \bibinfo {year} {1995})\BibitemShut {NoStop}%
\bibitem [{\citenamefont {Patel}\ \emph {et~al.}(2015)\citenamefont {Patel},
  \citenamefont {Strack},\ and\ \citenamefont {Sachdev}}]{subir-aavishkar}%
  \BibitemOpen
  \bibfield  {author} {\bibinfo {author} {\bibfnamefont {A.~A.}\ \bibnamefont
  {Patel}}, \bibinfo {author} {\bibfnamefont {P.}~\bibnamefont {Strack}},\ and\
  \bibinfo {author} {\bibfnamefont {S.}~\bibnamefont {Sachdev}},\ }\bibfield
  {title} {\bibinfo {title} {Hyperscaling at the spin density wave quantum
  critical point in two-dimensional metals},\ }\href
  {https://doi.org/10.1103/PhysRevB.92.165105} {\bibfield  {journal} {\bibinfo
  {journal} {Phys. Rev. B}\ }\textbf {\bibinfo {volume} {92}},\ \bibinfo
  {pages} {165105} (\bibinfo {year} {2015})}\BibitemShut {NoStop}%
\bibitem [{\citenamefont {Broerman}(1970)}]{Broerman_1}%
  \BibitemOpen
  \bibfield  {author} {\bibinfo {author} {\bibfnamefont {J.~G.}\ \bibnamefont
  {Broerman}},\ }\bibfield  {title} {\bibinfo {title} {Temperature dependence
  of the static dielectric constant of a symmetry-induced zero-gap
  semiconductor},\ }\href {https://doi.org/10.1103/PhysRevLett.25.1658}
  {\bibfield  {journal} {\bibinfo  {journal} {Phys. Rev. Lett.}\ }\textbf
  {\bibinfo {volume} {25}},\ \bibinfo {pages} {1658} (\bibinfo {year}
  {1970})}\BibitemShut {NoStop}%
\bibitem [{\citenamefont {Broerman}(1972)}]{Broerman_2}%
  \BibitemOpen
  \bibfield  {author} {\bibinfo {author} {\bibfnamefont {J.~G.}\ \bibnamefont
  {Broerman}},\ }\bibfield  {title} {\bibinfo {title}
  {Random-phase-approximation dielectric function of $a$-sn in the far
  infrared},\ }\href {https://doi.org/10.1103/PhysRevB.5.397} {\bibfield
  {journal} {\bibinfo  {journal} {Phys. Rev. B}\ }\textbf {\bibinfo {volume}
  {5}},\ \bibinfo {pages} {397} (\bibinfo {year} {1972})}\BibitemShut {NoStop}%
\bibitem [{\citenamefont {Boettcher}(2019)}]{Boettcher_PRB_2019}%
  \BibitemOpen
  \bibfield  {author} {\bibinfo {author} {\bibfnamefont {I.}~\bibnamefont
  {Boettcher}},\ }\bibfield  {title} {\bibinfo {title} {Optical response of
  luttinger semimetals in the normal and superconducting states},\ }\href
  {https://doi.org/10.1103/PhysRevB.99.125146} {\bibfield  {journal} {\bibinfo
  {journal} {Phys. Rev. B}\ }\textbf {\bibinfo {volume} {99}},\ \bibinfo
  {pages} {125146} (\bibinfo {year} {2019})}\BibitemShut {NoStop}%
\bibitem [{\citenamefont {Tchoumakov}\ and\ \citenamefont
  {Witczak-Krempa}(2019)}]{Witczak-Krempa_PRB_2019}%
  \BibitemOpen
  \bibfield  {author} {\bibinfo {author} {\bibfnamefont {S.}~\bibnamefont
  {Tchoumakov}}\ and\ \bibinfo {author} {\bibfnamefont {W.}~\bibnamefont
  {Witczak-Krempa}},\ }\bibfield  {title} {\bibinfo {title} {Dielectric and
  electronic properties of three-dimensional luttinger semimetals with a
  quadratic band touching},\ }\href
  {https://doi.org/10.1103/PhysRevB.100.075104} {\bibfield  {journal} {\bibinfo
   {journal} {Phys. Rev. B}\ }\textbf {\bibinfo {volume} {100}},\ \bibinfo
  {pages} {075104} (\bibinfo {year} {2019})}\BibitemShut {NoStop}%
\bibitem [{\citenamefont {Mauri}\ and\ \citenamefont
  {Polini}(2019)}]{Polini_PRB_2019}%
  \BibitemOpen
  \bibfield  {author} {\bibinfo {author} {\bibfnamefont {A.}~\bibnamefont
  {Mauri}}\ and\ \bibinfo {author} {\bibfnamefont {M.}~\bibnamefont {Polini}},\
  }\bibfield  {title} {\bibinfo {title} {Dielectric function and plasmons of
  doped three-dimensional luttinger semimetals},\ }\href
  {https://doi.org/10.1103/PhysRevB.100.165115} {\bibfield  {journal} {\bibinfo
   {journal} {Phys. Rev. B}\ }\textbf {\bibinfo {volume} {100}},\ \bibinfo
  {pages} {165115} (\bibinfo {year} {2019})}\BibitemShut {NoStop}%
\bibitem [{\citenamefont {Forster}(1975)}]{Forster-HFBSCF}%
  \BibitemOpen
  \bibfield  {author} {\bibinfo {author} {\bibfnamefont {D.}~\bibnamefont
  {Forster}},\ }\href@noop {} {\emph {\bibinfo {title} {{Hydrodynamic
  Fluctuations, Broken Symmetry, and Correlation Functions}}}}\ (\bibinfo
  {publisher} {W. A. Benjamin},\ \bibinfo {address} {Reading},\ \bibinfo {year}
  {1975})\BibitemShut {NoStop}%
\bibitem [{\citenamefont {Rosch}\ and\ \citenamefont
  {Andrei}(2000)}]{Rosch-PRL}%
  \BibitemOpen
  \bibfield  {author} {\bibinfo {author} {\bibfnamefont {A.}~\bibnamefont
  {Rosch}}\ and\ \bibinfo {author} {\bibfnamefont {N.}~\bibnamefont {Andrei}},\
  }\bibfield  {title} {\bibinfo {title} {{Conductivity of a Clean
  One-Dimensional Wire}},\ }\href {https://doi.org/10.1103/PhysRevLett.85.1092}
  {\bibfield  {journal} {\bibinfo  {journal} {Phys. Rev. Lett.}\ }\textbf
  {\bibinfo {volume} {85}},\ \bibinfo {pages} {1092} (\bibinfo {year}
  {2000})}\BibitemShut {NoStop}%
\bibitem [{\citenamefont {Mahajan}\ \emph {et~al.}(2013)\citenamefont
  {Mahajan}, \citenamefont {Barkeshli},\ and\ \citenamefont
  {Hartnoll}}]{Hartnoll-PRB_2013}%
  \BibitemOpen
  \bibfield  {author} {\bibinfo {author} {\bibfnamefont {R.}~\bibnamefont
  {Mahajan}}, \bibinfo {author} {\bibfnamefont {M.}~\bibnamefont {Barkeshli}},\
  and\ \bibinfo {author} {\bibfnamefont {S.~A.}\ \bibnamefont {Hartnoll}},\
  }\bibfield  {title} {\bibinfo {title} {Non-fermi liquids and the
  wiedemann-franz law},\ }\href {https://doi.org/10.1103/PhysRevB.88.125107}
  {\bibfield  {journal} {\bibinfo  {journal} {Phys. Rev. B}\ }\textbf {\bibinfo
  {volume} {88}},\ \bibinfo {pages} {125107} (\bibinfo {year}
  {2013})}\BibitemShut {NoStop}%
\bibitem [{\citenamefont {Freire}(2014)}]{Freire-AP_2014}%
  \BibitemOpen
  \bibfield  {author} {\bibinfo {author} {\bibfnamefont {H.}~\bibnamefont
  {Freire}},\ }\bibfield  {title} {\bibinfo {title} {{Controlled calculation of
  the thermal conductivity for a spinon Fermi surface coupled to a U(1) gauge
  field}},\ }\href {https://doi.org/10.1016/j.aop.2014.07.002} {\bibfield
  {journal} {\bibinfo  {journal} {Ann. Phys. (N. Y.)}\ }\textbf {\bibinfo
  {volume} {349}},\ \bibinfo {pages} {357} (\bibinfo {year}
  {2014})}\BibitemShut {NoStop}%
\bibitem [{\citenamefont {Patel}\ and\ \citenamefont
  {Sachdev}(2014)}]{Patel-PRB}%
  \BibitemOpen
  \bibfield  {author} {\bibinfo {author} {\bibfnamefont {A.~A.}\ \bibnamefont
  {Patel}}\ and\ \bibinfo {author} {\bibfnamefont {S.}~\bibnamefont
  {Sachdev}},\ }\bibfield  {title} {\bibinfo {title} {{dc resistivity at the
  onset of spin density wave order in two-dimensional metals}},\ }\href
  {https://doi.org/10.1103/PhysRevB.90.165146} {\bibfield  {journal} {\bibinfo
  {journal} {Phys. Rev. B}\ }\textbf {\bibinfo {volume} {90}},\ \bibinfo
  {pages} {165146} (\bibinfo {year} {2014})}\BibitemShut {NoStop}%
\bibitem [{\citenamefont {Hartnoll}\ \emph {et~al.}(2014)\citenamefont
  {Hartnoll}, \citenamefont {Mahajan}, \citenamefont {Punk},\ and\
  \citenamefont {Sachdev}}]{Hartnoll_PRB_2014}%
  \BibitemOpen
  \bibfield  {author} {\bibinfo {author} {\bibfnamefont {S.~A.}\ \bibnamefont
  {Hartnoll}}, \bibinfo {author} {\bibfnamefont {R.}~\bibnamefont {Mahajan}},
  \bibinfo {author} {\bibfnamefont {M.}~\bibnamefont {Punk}},\ and\ \bibinfo
  {author} {\bibfnamefont {S.}~\bibnamefont {Sachdev}},\ }\bibfield  {title}
  {\bibinfo {title} {Transport near the ising-nematic quantum critical point of
  metals in two dimensions},\ }\href
  {https://doi.org/10.1103/PhysRevB.89.155130} {\bibfield  {journal} {\bibinfo
  {journal} {Phys. Rev. B}\ }\textbf {\bibinfo {volume} {89}},\ \bibinfo
  {pages} {155130} (\bibinfo {year} {2014})}\BibitemShut {NoStop}%
\bibitem [{\citenamefont {Zaanen}\ \emph {et~al.}(2015)\citenamefont {Zaanen},
  \citenamefont {Liu}, \citenamefont {Sun},\ and\ \citenamefont
  {Schalm}}]{Zaanen-CUP}%
  \BibitemOpen
  \bibfield  {author} {\bibinfo {author} {\bibfnamefont {J.}~\bibnamefont
  {Zaanen}}, \bibinfo {author} {\bibfnamefont {Y.}~\bibnamefont {Liu}},
  \bibinfo {author} {\bibfnamefont {Y.-W.}\ \bibnamefont {Sun}},\ and\ \bibinfo
  {author} {\bibfnamefont {K.}~\bibnamefont {Schalm}},\ }\href@noop {} {\emph
  {\bibinfo {title} {{Holographic Duality in Condensed Matter Physics}}}}\
  (\bibinfo  {publisher} {Cambridge University Press},\ \bibinfo {address}
  {Cambridge},\ \bibinfo {year} {2015})\BibitemShut {NoStop}%
\bibitem [{\citenamefont {Freire}(2017{\natexlab{a}})}]{Freire-AP_2017}%
  \BibitemOpen
  \bibfield  {author} {\bibinfo {author} {\bibfnamefont {H.}~\bibnamefont
  {Freire}},\ }\bibfield  {title} {\bibinfo {title} {{Memory matrix theory of
  the dc resistivity of a disordered antiferromagnetic metal with an effective
  composite operator}},\ }\href {https://doi.org/10.1016/j.aop.2017.07.001}
  {\bibfield  {journal} {\bibinfo  {journal} {Ann. Phys. (N. Y.)}\ }\textbf
  {\bibinfo {volume} {384}},\ \bibinfo {pages} {142} (\bibinfo {year}
  {2017}{\natexlab{a}})}\BibitemShut {NoStop}%
\bibitem [{\citenamefont {Freire}(2017{\natexlab{b}})}]{Freire-EPL}%
  \BibitemOpen
  \bibfield  {author} {\bibinfo {author} {\bibfnamefont {H.}~\bibnamefont
  {Freire}},\ }\bibfield  {title} {\bibinfo {title} {{Calculation of the
  magnetotransport for a spin-density-wave quantum critical theory in the
  presence of weak disorder}},\ }\href
  {https://doi.org/10.1209/0295-5075/118/57003} {\bibfield  {journal} {\bibinfo
   {journal} {{EPL} (Europhysics Letters)}\ }\textbf {\bibinfo {volume}
  {118}},\ \bibinfo {pages} {57003} (\bibinfo {year}
  {2017}{\natexlab{b}})}\BibitemShut {NoStop}%
\bibitem [{\citenamefont {Hartnoll}\ \emph {et~al.}(2018)\citenamefont
  {Hartnoll}, \citenamefont {Lucas},\ and\ \citenamefont
  {Sachdev}}]{Sachdev-MIT}%
  \BibitemOpen
  \bibfield  {author} {\bibinfo {author} {\bibfnamefont {S.~A.}\ \bibnamefont
  {Hartnoll}}, \bibinfo {author} {\bibfnamefont {A.}~\bibnamefont {Lucas}},\
  and\ \bibinfo {author} {\bibfnamefont {S.}~\bibnamefont {Sachdev}},\
  }\href@noop {} {\emph {\bibinfo {title} {{Holographic Quantum Matter}}}}\
  (\bibinfo  {publisher} {MIT Press},\ \bibinfo {address} {Cambridge},\
  \bibinfo {year} {2018})\BibitemShut {NoStop}%
\bibitem [{\citenamefont {Freire}(2018)}]{Freire-EPL_2018}%
  \BibitemOpen
  \bibfield  {author} {\bibinfo {author} {\bibfnamefont {H.}~\bibnamefont
  {Freire}},\ }\bibfield  {title} {\bibinfo {title} {{Thermal and
  thermoelectric transport coefficients for a two-dimensional {SDW} metal with
  weak disorder: A memory matrix calculation}},\ }\href
  {https://doi.org/10.1209/0295-5075/124/27003} {\bibfield  {journal} {\bibinfo
   {journal} {{EPL} (Europhysics Letters)}\ }\textbf {\bibinfo {volume}
  {124}},\ \bibinfo {pages} {27003} (\bibinfo {year} {2018})}\BibitemShut
  {NoStop}%
\bibitem [{\citenamefont {Wang}\ and\ \citenamefont {Berg}(2019)}]{Berg-PRB}%
  \BibitemOpen
  \bibfield  {author} {\bibinfo {author} {\bibfnamefont {X.}~\bibnamefont
  {Wang}}\ and\ \bibinfo {author} {\bibfnamefont {E.}~\bibnamefont {Berg}},\
  }\bibfield  {title} {\bibinfo {title} {{Scattering mechanisms and electrical
  transport near an Ising nematic quantum critical point}},\ }\href
  {https://doi.org/10.1103/PhysRevB.99.235136} {\bibfield  {journal} {\bibinfo
  {journal} {Phys. Rev. B}\ }\textbf {\bibinfo {volume} {99}},\ \bibinfo
  {pages} {235136} (\bibinfo {year} {2019})}\BibitemShut {NoStop}%
\bibitem [{\citenamefont {Vieira}\ \emph {et~al.}(2020)\citenamefont {Vieira},
  \citenamefont {{de Carvalho}},\ and\ \citenamefont
  {Freire}}]{Freire-AP_2020}%
  \BibitemOpen
  \bibfield  {author} {\bibinfo {author} {\bibfnamefont {L.~E.}\ \bibnamefont
  {Vieira}}, \bibinfo {author} {\bibfnamefont {V.~S.}\ \bibnamefont {{de
  Carvalho}}},\ and\ \bibinfo {author} {\bibfnamefont {H.}~\bibnamefont
  {Freire}},\ }\bibfield  {title} {\bibinfo {title} {Dc resistivity near a
  nematic quantum critical point: Effects of weak disorder and acoustic
  phonons},\ }\href {https://doi.org/https://doi.org/10.1016/j.aop.2020.168230}
  {\bibfield  {journal} {\bibinfo  {journal} {Annals of Physics}\ }\textbf
  {\bibinfo {volume} {419}},\ \bibinfo {pages} {168230} (\bibinfo {year}
  {2020})}\BibitemShut {NoStop}%
\bibitem [{\citenamefont {Wang}\ and\ \citenamefont
  {Berg}(2020)}]{wang2020low}%
  \BibitemOpen
  \bibfield  {author} {\bibinfo {author} {\bibfnamefont {X.}~\bibnamefont
  {Wang}}\ and\ \bibinfo {author} {\bibfnamefont {E.}~\bibnamefont {Berg}},\
  }\href@noop {} {\bibinfo {title} {Low frequency raman response near
  ising-nematic quantum critical point: a memory matrix approach}} (\bibinfo
  {year} {2020}),\ \Eprint {https://arxiv.org/abs/2011.01818} {arXiv:2011.01818
  [cond-mat.str-el]} \BibitemShut {NoStop}%
\bibitem [{\citenamefont {Kumar}\ \emph {et~al.}(2020)\citenamefont {Kumar},
  \citenamefont {Kharkwal}, \citenamefont {Kumar}, \citenamefont {Asokan},
  \citenamefont {Banerjee},\ and\ \citenamefont {Pramanik}}]{Pramanik}%
  \BibitemOpen
  \bibfield  {author} {\bibinfo {author} {\bibfnamefont {H.}~\bibnamefont
  {Kumar}}, \bibinfo {author} {\bibfnamefont {K.~C.}\ \bibnamefont {Kharkwal}},
  \bibinfo {author} {\bibfnamefont {K.}~\bibnamefont {Kumar}}, \bibinfo
  {author} {\bibfnamefont {K.}~\bibnamefont {Asokan}}, \bibinfo {author}
  {\bibfnamefont {A.}~\bibnamefont {Banerjee}},\ and\ \bibinfo {author}
  {\bibfnamefont {A.~K.}\ \bibnamefont {Pramanik}},\ }\bibfield  {title}
  {\bibinfo {title} {Magnetic and transport properties of the pyrochlore
  iridates
  $({\mathrm{y}}_{1\ensuremath{-}x}{\mathrm{pr}}_{x}{)}_{2}{\mathrm{ir}}_{2}{\mathrm{o}}_{7}$:
  Role of $f\text{\ensuremath{-}}d$ exchange interaction and
  $d\text{\ensuremath{-}}p$ orbital hybridization},\ }\href
  {https://doi.org/10.1103/PhysRevB.101.064405} {\bibfield  {journal} {\bibinfo
   {journal} {Phys. Rev. B}\ }\textbf {\bibinfo {volume} {101}},\ \bibinfo
  {pages} {064405} (\bibinfo {year} {2020})}\BibitemShut {NoStop}%
\bibitem [{\citenamefont {Cheng}\ \emph {et~al.}(2017)\citenamefont {Cheng},
  \citenamefont {Ohtsuki}, \citenamefont {Chaudhuri}, \citenamefont
  {Nakatsuji}, \citenamefont {Lippmaa},\ and\ \citenamefont
  {Armitage}}]{armitage_expt}%
  \BibitemOpen
  \bibfield  {author} {\bibinfo {author} {\bibfnamefont {B.}~\bibnamefont
  {Cheng}}, \bibinfo {author} {\bibfnamefont {T.}~\bibnamefont {Ohtsuki}},
  \bibinfo {author} {\bibfnamefont {D.}~\bibnamefont {Chaudhuri}}, \bibinfo
  {author} {\bibfnamefont {S.}~\bibnamefont {Nakatsuji}}, \bibinfo {author}
  {\bibfnamefont {M.}~\bibnamefont {Lippmaa}},\ and\ \bibinfo {author}
  {\bibfnamefont {N.~P.}\ \bibnamefont {Armitage}},\ }\bibfield  {title}
  {\bibinfo {title} {Dielectric anomalies and interactions in the
  three-dimensional quadratic band touching luttinger semimetal pr2ir2o7},\
  }\href {https://doi.org/10.1038/s41467-017-02121-y} {\bibfield  {journal}
  {\bibinfo  {journal} {Nature Communications}\ }\textbf {\bibinfo {volume}
  {8}},\ \bibinfo {pages} {2097} (\bibinfo {year} {2017})}\BibitemShut
  {NoStop}%
\bibitem [{\citenamefont {{Freire}}\ and\ \citenamefont
  {{Mandal}}(2021)}]{ips-hermann2}%
  \BibitemOpen
  \bibfield  {author} {\bibinfo {author} {\bibfnamefont {H.}~\bibnamefont
  {{Freire}}}\ and\ \bibinfo {author} {\bibfnamefont {I.}~\bibnamefont
  {{Mandal}}},\ }\bibfield  {title} {\bibinfo {title} {{Thermoelectric and
  thermal properties of the weakly disordered non-Fermi liquid phase of
  Luttinger semimetals}},\ }\href@noop {} {\bibfield  {journal} {\bibinfo
  {journal} {arXiv e-prints}\ } (\bibinfo {year} {2021})},\ \Eprint
  {https://arxiv.org/abs/2104.07459} {arXiv:2104.07459 [cond-mat.str-el]}
  \BibitemShut {NoStop}%
\bibitem [{\citenamefont {Janssen}\ and\ \citenamefont
  {Herbut}(2015)}]{lukas-herbut}%
  \BibitemOpen
  \bibfield  {author} {\bibinfo {author} {\bibfnamefont {L.}~\bibnamefont
  {Janssen}}\ and\ \bibinfo {author} {\bibfnamefont {I.~F.}\ \bibnamefont
  {Herbut}},\ }\bibfield  {title} {\bibinfo {title} {Nematic quantum
  criticality in three-dimensional fermi system with quadratic band touching},\
  }\href {https://doi.org/10.1103/PhysRevB.92.045117} {\bibfield  {journal}
  {\bibinfo  {journal} {Phys. Rev. B}\ }\textbf {\bibinfo {volume} {92}},\
  \bibinfo {pages} {045117} (\bibinfo {year} {2015})}\BibitemShut {NoStop}%
\end{thebibliography}%

\appendix

%%%%%%%%%%%%%%%%%%%%%%%%%%%%%%%%%%%%%%%%%%%%%%%%

\begin{widetext}
%%%%%%%%%%%%%%%%%%%%%%%%%%%%%%%%%%%%%%%%%%%

\section{$d_a$-function algebra}
\label{angular}

We derive a set of useful relations \cite{lukas-herbut,igor16} for the vector functions $\mathbf{d}(\mathbf{k})$ (whose components $d_a(\mathbf{k})$ are the $\ell=2$ spherical harmonics in $d$ spatial dimensions) and the generalized real $d \times d$ Gell-Mann matrices $\Lambda_a$ ($a = 1,2,\cdots, N$).
The  matrices $\Lambda_a$ are  symmetric,  traceless, and orthogonal, satisfying
\begin{align}
& \text{Tr}[  \Lambda^a \,  \Lambda^b] =2\,\delta_{ab} \,,
\quad \sum_{a=1}^N \left ( \Lambda^a \right )_{ij} \left ( \Lambda^a_{l j'}\right )
=\delta_{i l} \, \delta_{j j'}+ \delta_{i j'}\, \delta_{jl}
- \frac{2} {d}\,\delta_{ i j }\, \delta_{l j'}\,.
\end{align}
Hence, the index $a$ (or $b$) runs from $1$ to $N = \frac{\left (d-1\right)\left (d+2\right)}{2}$.
We define the components of $\mathbf{d}(\mathbf k)$ by
\begin{align}
d_a(\mathbf{k}) =\sqrt { \frac{d}{2 \left(d-1 \right ) }}
\sum \limits_{i,j=1}^{d}
\frac{ k_i \left ( \Lambda^a\right)_{ij}  k _j } {2\,m}\,.
\end{align}
This gives the following identities:
\begin{align}
&\partial_{k_{z}}d_a(\mathbf{k}) = 
\sqrt { \frac{d}{2 \left(d-1 \right ) }}\,
\frac{\sum \limits_{j=1}^{d}  \left ( \Lambda^a\right)_{zj}  k _j
+ \sum \limits_{i=1}^{d} k_i \left ( \Lambda^a\right)_{ i z}  } {2\,m}
=
\sqrt { \frac{2\,d}{ d-1  }} \, \frac{
 \sum \limits_{j=1}^{d}  \left ( \Lambda^a\right)_{zj} k _j } {2\,m}\,,
%%%%%%%%55
\nn & \sum \limits_{a=1}^N \left \lbrace \partial_{k_z} d_a(\mathbf{k})  \right \rbrace^2
 = \frac{2\,d} { d-1 } \sum \limits_{i}^{d}
 \frac{ \left ( 
\delta_{i i} + \delta_{i z}\, \delta_{z i}
- \frac{2} {d}\,\delta_{ i z }\, \delta_{ i z} 
  \right)  k_i^2 } {4 \,m^2}
%%%%
= \frac{d \times \left ( 
\mathbf  k^2 +  \frac{d-2} {d}\,k_z^2 \right)} { 2\,m^2 \left( d-1 \right)}   \,,
\nn
& \sum \limits _{a=1}^N 
 d_a(\mathbf{k}) \, d_a(\mathbf{p}) = 
 \frac{  d 
 \left (\mathbf{k}\cdot\mathbf{p}  \right )^2 - \mathbf  k^2\,\mathbf  p^2 } 
 {4\,m^2 \left( d-1\right) } \,,
\quad
\sum \limits _{a=1}^N 
\left[  \partial_{k_z} d_a(\mathbf{k}) \right ] d_a(\mathbf{p}) 
= \frac{  k_z \left \lbrace d \left (\mathbf{k}\cdot\mathbf{p}  \right )
-   \mathbf  p^2 \right \rbrace }
{ 2\,m^2 \left(d-1 \right)} \,,
 \nn &
  \sum \limits _{a=1}^N 
\left[  \partial_{p_z} d_a(\mathbf{p}) \right ] 
\left[  \partial_{k_z} d_a(\mathbf{k}) \right ]
 = \sum_{a=1}^N \frac{2\,d}{ 4\,m^2 \left(d-1 \right) } 
 \sum \limits_{j=1}^{d}  \left ( \Lambda^a\right)_{zj}   p_j 
 \,  \sum \limits_{j'=1}^{d}  \left ( \Lambda^a\right)_{zj'}   k_{j'}  
 =
\frac{   d \, \mathbf{p} \cdot \mathbf{k }
+   (d-2)\, p_z\,k _{z}  } {2\,m^2 \left(d-1 \right)} \,.
 \label{eqdprod3}
\end{align}

For the special case of $\mathbf{p}=\mathbf{k}$, we obtain:
\begin{align}
 \label{d5}  \sum_{a=1}^N  d_a^2(\mathbf{k})
  = \frac{ \mathbf k^4}{4\,m^2}\,, \quad
\frac{1}{2}\partial_{k_z}\sum_{a=1}^N  d_a^2(\mathbf{k})
=\frac{ k_z \mathbf k^2 }{ 2\,m^2}  \,  .
\end{align}
%For $d=4$, we get
%\begin{align}
%\label{d4} 
%4\,m^2 \sum_{a=1}^4  d_a(\mathbf{p})\, d_a(\mathbf{k}) =\frac{1}{3}
%\left [ 4\,(\mathbf{p}\cdot\mathbf{k})^2 - \mathbf p^2\,\mathbf  k^2 \right  ] .
%\end{align}

%%%%%%%%%%%%%%%%%%%%%%%%%%%%%%%%%%%%
 \section{Two-Loop Contributions to the current-current correlators}
\label{current2loop}

\subsection{Self-energy corrections}
\label{app2loop1}

The diagrams in Figs.~\ref{fig2} and \ref{fig3} involve inserting the one-loop fermion self-energy ($\Sigma_1$) corrections into the current-current correlator.
We include a factor of $2$ as the two diagrams give equal contributions and the expression incorporating this correction takes the form
%%%%%%%%%%%%%%%%%%%%%%
\begin{align}
& \langle J_z J_z \rangle_\text{2loop}^{(1)}(\mathrm{i}\,\omega)
= - 2  \int \frac{dk_0}{2\pi} 
\int \frac{d^d {\mathbf k}}{(2\pi)^{d}}
\text{Tr} \left [ \left \lbrace 
\partial_{k_z}\mathbf{d}(\mathbf k )\cdot \mathbf \Gamma\right \rbrace
G_0(k+q) \,
\Sigma_1(k+q)
G_0(k+q)\left \lbrace 
\partial_{k_z}\mathbf{d}(\mathbf k)\cdot \mathbf \Gamma\right \rbrace 
 \,G_0(k)
\right ]\,,
%%%%%%%%%%%%%%
\end{align}
where $\Sigma_1 (\boldsymbol \ell) 
= -\frac{ m\,e^2  } 
{15\,\pi^2\,c} \left( \frac{\Lambda^{1/2}}{|\boldsymbol \ell|} \right)^\varepsilon  
\frac{ \mathbf{d}(\boldsymbol \ell) \cdot \mathbf{\Gamma} }
{\varepsilon}$
(see Refs. \cite{rahul-sid,ips-rahul,*ips-rahul-errata}).
This gives us:
\begin{align}
& \langle J_z J_z \rangle_\text{2loop}^{(1)}(\mathrm{i}\,\omega)
= \frac{ 2\, m\,e^2 } 
{15\,\pi^2\,c\,\varepsilon }   
 \int \frac{dk_0}{2\pi} 
\int \frac{d^d {\mathbf k}}{(2\pi)^{d}} \left( \frac{\Lambda^{1/2}}{|\mathbf k|} \right)^\varepsilon
\frac{term_1
}
{
 \left\lbrace k_0 ^2 +|\mathbf{d}(\mathbf{k})|^2 \right \rbrace^2
\left \lbrace \left (  k_0+ \omega  \right )^2 +|\mathbf{d}(\mathbf{k})|^2 \right \rbrace
}
\,,
%%%%%%%%%%%%%%
\end{align}
where
\begin{align}
term_1
&=\text{Tr} \left [
\left \lbrace 
\partial_{k_z}\mathbf{d}(\mathbf k )\cdot \mathbf \Gamma\right \rbrace
{ \left\lbrace
\mathrm{i}\, k_0   + \mathbf{d}(\mathbf{k} ) \cdot{\mathbf{\Gamma}}
\right \rbrace }
 \left \lbrace \mathbf{d}(\mathbf k) \cdot \mathbf{\Gamma}\right \rbrace
{ \left\lbrace
\mathrm{i}\, k_0   + \mathbf{d}(\mathbf{k} ) \cdot{\mathbf{\Gamma}}\right \rbrace}
\left \lbrace 
\partial_{k_z}\mathbf{d}(\mathbf k)\cdot \mathbf \Gamma\right \rbrace 
{ \left \lbrace \mathrm{i}\, k_0 + \mathrm{i}\,\omega  + \mathbf{d}(\mathbf{k} ) \cdot{\mathbf{\Gamma}} \right \rbrace} \right ]\nn
%%%%%%%%%%%%%%%%%%
&=-k_0^2 \,\text{Tr} \left [
\left \lbrace 
\partial_{k_z}\mathbf{d}(\mathbf k )\cdot \mathbf \Gamma\right \rbrace 
 \left \lbrace \mathbf{d}(\mathbf k) \cdot \mathbf{\Gamma}\right \rbrace  
\left \lbrace 
\partial_{k_z}\mathbf{d}(\mathbf k)\cdot \mathbf \Gamma\right \rbrace 
{ \left \lbrace    \mathbf{d}(\mathbf{k} ) \cdot{\mathbf{\Gamma}} \right \rbrace} \right ]\nn
& \quad 
-2\,k_0 \left( k_0 +\omega \right)
\text{Tr} \left [
\left \lbrace 
\partial_{k_z}\mathbf{d}(\mathbf k )\cdot \mathbf \Gamma\right \rbrace
 \left \lbrace \mathbf{d}(\mathbf k) \cdot \mathbf{\Gamma}\right \rbrace
{ \left\lbrace \mathbf{d}(\mathbf{k} ) \cdot{\mathbf{\Gamma}}\right \rbrace}
\left \lbrace 
\partial_{k_z}\mathbf{d}(\mathbf k)\cdot \mathbf \Gamma\right \rbrace 
 \right ]
\nn
& \quad 
+
\text{Tr} \left [
\left \lbrace 
\partial_{k_z}\mathbf{d}(\mathbf k )\cdot \mathbf \Gamma\right \rbrace
{ \left\lbrace \mathbf{d}(\mathbf{k} ) \cdot{\mathbf{\Gamma}}
\right \rbrace }
 \left \lbrace \mathbf{d}(\mathbf k) \cdot \mathbf{\Gamma}\right \rbrace
{ \left\lbrace
 \mathbf{d}(\mathbf{k} ) \cdot{\mathbf{\Gamma}}\right \rbrace}
\left \lbrace 
\partial_{k_z}\mathbf{d}(\mathbf k)\cdot \mathbf \Gamma\right \rbrace 
{ \left \lbrace  \mathbf{d}(\mathbf{k} ) \cdot{\mathbf{\Gamma}} \right \rbrace} \right ]\nn
%%%%%%%%%%%%%%%%%%%%%%%%%%%%%%%%%%%
&=-4\,k_0^2 \left [
2 \left \lbrace 
\partial_{k_z}\mathbf{d}(\mathbf k )\cdot 
 \mathbf{d}(\mathbf k) \right \rbrace ^2
-\left \lbrace 
\partial_{k_z}\mathbf{d}(\mathbf k )\cdot 
\partial_{k_z} \mathbf{d}(\mathbf k) \right \rbrace
| \mathbf{d}(\mathbf k )|^2 
    \right ]
%%%%%%%%
  -8\,k_0 \left( k_0 +\omega \right)
\left \lbrace 
\partial_{k_z}\mathbf{d}(\mathbf k )\cdot 
\partial_{k_z} \mathbf{d}(\mathbf k) \right \rbrace
| \mathbf{d}(\mathbf k )|^2  \nn
%%%%%%%%%%%%%%%%%%%%%%%%%%%%%%%%%%%%%%%%55
& \quad
+ 8  \left \lbrace 
\partial_{k_z}\mathbf{d}(\mathbf k )\cdot 
 \mathbf{d}(\mathbf k) \right \rbrace^2
 | \mathbf{d}(\mathbf k )|^2 
-4  \left \lbrace 
\partial_{k_z}\mathbf{d}(\mathbf k )\cdot 
\partial_{k_z} \mathbf{d}(\mathbf k) \right  \rbrace | \mathbf{d}(\mathbf k )|^4
\nn
%%%%%%%%%%%%%%%%%%%%%%%%%%%%%%%%%%%%%%%%
& =k_0^2 \,\mathbf k^6 \times
\frac{ \left (6-5 \,d \right) \sin ^2  \theta -d}{2 \left(d-1 \right) m^4}
-k_0 \,\omega\,\mathbf k^6 \times
\frac{
d +  \left( d-2\right)\sin^2 \theta  } 
{   \left( d-1 \right) m^4}
+ \frac{   \mathbf k^{10}
\left [ 
  \left( 3\,d-2 \right) \sin^2 \theta-d \right ]} {8\left( d-1 \right) m^6} \,,
\end{align}
%https://arxiv.org/pdf/1710.05164.pdf eqn38
using the identities from Appendix~\ref{angular}.
 
Performing the integrals, we finally get:
\begin{align}
\langle J_z J_z \rangle_\text{2loop}^{(1)}(\mathrm{i}\,\omega) 
= \frac{e^2 \,  m^{2-\frac{\varepsilon }{2}} \,| \omega | ^{2-\frac{\varepsilon }{2}}}
{90\, \pi ^4 \,c \,\varepsilon ^2} 
\left( \frac{\Lambda}{m\,\omega} \right)^{\varepsilon/2}
-\frac{e^2 \,  m^{2-\frac{\varepsilon }{2}} | \omega | ^{2-\frac{\varepsilon }{2}} 
\ln\left(\frac{m \,| \omega | }{\Lambda }\right)}
{180\, \pi ^4 \,c\, \varepsilon }\,.
 \end{align}

%%%%%%%%%%%%%%%%%%%%%%%%%%%%%%%%%
\subsection{Vertex corrections}
\label{app2loop2} 

The diagram in Fig.~\ref{fig4} equals $ \langle J_z J_z \rangle_\text{2loop}^{(2)}(\mathrm{i}\,\omega)$, where
%%%%%%%%%%%%%%%%%%%%%%%%%%%%%%%%%%%%%%%%%%
\begin{align}
& \frac{\langle J_z J_z \rangle_\text{2loop}^{(2)}(\mathrm{i}\,\omega)}
{\frac{e^2\,\Lambda^{\varepsilon/2}}{c}}
\nn
& =  \int \frac{dk_0\,d\ell_0}  {(2\pi)^2 }
\int \frac{d^d {\mathbf k}\, d^d {\boldsymbol{\ell}}}{(2\pi)^{2d}}
\text{Tr} \left [ \left \lbrace 
\partial_{{k_z}}
\mathbf{d}(\mathbf k )\cdot \mathbf \Gamma\right \rbrace
G_0(k+q) \,\frac{1} {\boldsymbol{\ell}^2}\, G_0(k+q+\ell)
\left \lbrace 
\partial_{k_z+\ell_z}
\mathbf{d}(\mathbf k + \boldsymbol{\ell})\cdot \mathbf \Gamma\right \rbrace G_0(k+\ell)
\,G_0(k)
\right ]\nn
%%%%%%%%%%%%%%
& =  
\int \frac{dk_0\,d\ell_0}  {(2\pi)^2 }
\int \frac{d^d {\mathbf k}\, d^d {\boldsymbol{\ell}}}{(2\pi)^{2d}}
\text{Tr} \left [ \left \lbrace 
\partial_{{k_z}}
\mathbf{d}(\mathbf k )\cdot \mathbf \Gamma\right \rbrace
G_0(k_0+\omega,\mathbf k) \,\frac{1} {\boldsymbol{\ell}^2}\, 
G_0( \ell_0 + \omega,\mathbf k+\boldsymbol \ell)
\left \lbrace 
\partial_{k_z+\ell_z}
\mathbf{d}(\mathbf k + \boldsymbol{\ell})\cdot \mathbf \Gamma\right \rbrace 
G_0(\ell_0,\mathbf k +\boldsymbol \ell)
\,G_0(k_0,\mathbf k)
\right ],
%%%%%%%%%%%%%%
\end{align}
with $\ell =(\ell_0, \boldsymbol{ \ell } )$.
%%%%%%%%%%%%%%
 We observe that the expression to be evaluated is
%%%%%%%%%%%%%%%%%%%%%%%%%%%%%%%%%%%%%%%%%%
\begin{align}
& \frac{\langle J_z J_z \rangle_\text{2loop}^{(2)}(\mathrm{i}\,\omega)}
{\frac{e^2\,\Lambda^{\varepsilon/2}} {c}}
 = 
\int \frac{d^d {\mathbf k}\, d^d {\boldsymbol{\ell}}}{(2\pi)^{2d}}
\text{Tr} \left [ 
\frac{\int \frac{dk_0}  {2\pi }\,G_0(k_0-\omega,\mathbf k)
\left \lbrace 
\partial_{{k_z}}
\mathbf{d}(\mathbf k )\cdot \mathbf \Gamma\right \rbrace
G_0(k_0 ,\mathbf k) \,
%%%%%%%%%%%%% 
\int \frac{d\ell_0}  {2\pi }\,
G_0(\ell_0+ \omega, \boldsymbol \ell)
\left \lbrace 
\partial_{ \ell_z}
\mathbf{d}(  \boldsymbol{\ell})\cdot \mathbf \Gamma\right \rbrace 
G_0(\ell_0, \boldsymbol \ell)}
{\left( \mathbf k +  \boldsymbol{\ell} \right )^2}
\right ],
\end{align}
after some clever regrouping of the terms in the integrand.
Evaluating
\begin{align}
& \int \frac{d\ell_0}  {2\pi }\,
G_0(\ell_0+ \omega, \boldsymbol \ell)
\left \lbrace 
\partial_{ \ell_z}
\mathbf{d}(  \boldsymbol{\ell})\cdot \mathbf \Gamma\right \rbrace 
 G_0(\ell_0, \boldsymbol \ell)
=
\int \frac{d\ell_0}  {2\pi }\,
\frac{
- \ell_0 \left(\ell_0 + \omega \right) 
\partial_{\ell_z} \mathbf{d} (\boldsymbol \ell) \cdot \mathbf{\Gamma}  + 
 \frac{\mathrm{i} \left( 2\,\ell_0 + \omega \right) \ell_z\, \boldsymbol \ell^2} {2 \,m^2} 
 + \frac{\mathbf{d} (\boldsymbol \ell) \cdot \mathbf{\Gamma}
 \,\ell_z \,\boldsymbol \ell^2 } {2 \,m^2}
}
{\left \lbrace \left (  \ell_0+  \omega \right )^2
+| \mathbf{d}(  \boldsymbol{\ell})|^2
\right \rbrace
\left \lbrace
   \ell_0^2 + | \mathbf{d}(  \boldsymbol{\ell})|^2 \right \rbrace
}
%%%%%%%%%%%%
\nn &
= \frac{   \left[ 
2 \, {\ell}_z\, \mathbf{d}(  \boldsymbol{\ell}) 
-\boldsymbol{\ell}^2 \,\partial_{\ell_z} \mathbf{d}(  \boldsymbol{\ell})
\right ]
\cdot \mathbf \Gamma
}
{2 \,m
\left( \frac{\boldsymbol{\ell}^4} {m^2} +  \,\omega ^2\right)}\,,
\end{align}
%%%%%%%%%%%%%%%%%%%%%%%%%%%%%%%%%%%
we get:
%%%%%%%%%%%%%%%%%%%%%%%%%%%%%%%%%%%%%%%%%%
\begin{align}
& \frac{\langle J_z J_z \rangle_\text{2loop}^{(2)}(\mathrm{i}\,\omega)}
{\frac{e^2\,\Lambda^{\varepsilon/2}} {c}}
 =
\int \frac{d^d {\mathbf k}\, d^d {\boldsymbol{\ell}}}{(2\pi)^{2d}}
\text{Tr} \left [ 
\frac{
\left[ 2  \,k_z 
\left \lbrace \mathbf{d} (\mathbf k ) \cdot \mathbf{\Gamma} \right \rbrace    
-
\mathbf k^2\left \lbrace \partial_{k_z} \mathbf{d} (\mathbf k) \cdot \mathbf{\Gamma} \right \rbrace \right ]
%%%%%%%
\left[  2 \,\ell_z 
\left \lbrace \mathbf{d} (\boldsymbol \ell) \cdot \mathbf{\Gamma} \right \rbrace    
-
 \boldsymbol \ell^2 \left \lbrace \partial_{k_z} \mathbf{d} (\boldsymbol \ell) \cdot \mathbf{\Gamma} \right \rbrace \right ]
}
{ 4\,m^2 \left( \mathbf k +  \boldsymbol{\ell} \right )^2
 \left( \frac{\mathbf k^4} {m^2 } + \omega ^2\right) \left( \frac{\boldsymbol \ell^4} {m^2 } + \omega ^2\right)
}
\right ]
%%%%%%%%%%%%%%%%%%%%%%%%%%%%%%%%%%%%%%%%%%%%%%%%%%
\nn & =
 \int \frac{d^d {\mathbf k}\, d^d {\boldsymbol{\ell}}}{(2\pi)^{2d}}
\frac{
k_z \,\ell_z\,
\left \lbrace  d \,
 \left (\mathbf{k}\cdot \boldsymbol{\ell}  \right )^2 - \mathbf  k^2\,\boldsymbol{\ell}^2 
\right \rbrace
-2 \,k_z\,\,\ell_z\,\mathbf k^2 
  \left \lbrace d \left (\mathbf{k}\cdot \boldsymbol{\ell}  \right )
- \boldsymbol{\ell}^2 \right \rbrace 
%%%
 +
\mathbf k^2\,\boldsymbol \ell^2\,
\frac{ d \, \mathbf{k} \cdot  \boldsymbol \ell 
+   (d-2)\,k _{z}\, \ell_z} 
{ 2 }  
%%%
}
%%%%%%%
{ m^4 \left(d-1 \right) \left( \mathbf k +  \boldsymbol{\ell} \right )^2
 \left( \frac{\mathbf k^4} {m^2 } + \omega ^2\right) \left( \frac{\boldsymbol \ell^4} {m^2 } + \omega ^2\right)
}\,,
\end{align}
using the identities from Appendix~\ref{angular}.
Performing the integrals, we finally obtain:
\begin{align}
\langle J_z J_z \rangle_\text{2loop}^{(2)}(\mathrm{i}\,\omega) 
= \frac{ e^2\, 
m^{2-\frac{\varepsilon }{2}} \,| \omega | ^{2-\frac{\varepsilon }{2}}
\left( \frac{\Lambda}{m\,|\omega|} \right)^{\varepsilon/2}
}
{ 60 \,\pi ^4 \,c\, \varepsilon ^2}
-\frac{  e^2 \, m^{2-\frac{\varepsilon }{2}} \,| \omega | ^{2-\frac{\varepsilon }{2}} 
\ln\left(\frac{m \,| \omega | }{\Lambda }\right)}
{ 120 \,\pi ^4\, c \,\varepsilon }\,.
%%%%%
\end{align}

%%%%%%%%%%%%%%%%%%%%%%%%%%%%%%%%%%%%%%%%%%%%%%%%
\section{Two-loop contributions to the current-momentum susceptibility}
\label{chi2loop}

For the contributions at two-loop order represented by diagrams with self-energy insertions (similar to the diagrams in Figs.~\ref{fig2} and \ref{fig3}), we get the expression:
\begin{align}
\label{chiJP1}
& \chi^{(2,1)}_{J_z P_z}(T)= - 2\,T\, 
\sum \limits_{k_0}
\int \frac{d^3 {\mathbf k}}{(2\pi)^3}\,  
k_z \,
 \text{Tr}\left[ 
 \left \lbrace  \partial_{k_z}\mathbf{d}{(\mathbf k)}\cdot \mathbf \Gamma \right \rbrace
 G_0(k_0,\mathbf{k})
\,\Sigma_T(k_0,\mathbf{k})
\, G_0(k_0,\mathbf{k}) \,G_0(k_0,\mathbf{k})   \right],
\end{align}
where 
\begin{align}
\label{chiJP1-2}
\Sigma_T(k_0,\mathbf{k})
& =-\frac{e^2}{c}\,T
\sum \limits _{\ell_0}\int \frac{d^3 {\boldsymbol \ell}}{(2\pi)^3}
\frac{G_0(k_0+\ell_0,\mathbf k+\boldsymbol \ell)}
{\boldsymbol \ell^2}
=
-\frac{e^2}{15\pi^2 c}\left[\frac{\Lambda_0}{T}
\left \lbrace \mathbf{d}{(\mathbf k)}\cdot \mathbf \Gamma\right \rbrace
-\frac{5\,m}{4\,\Lambda_{IR}}\,\big|\mathbf{d}{(\mathbf k)} \big | \right],
\end{align}
where $\Lambda_0$ and $\Lambda_{IR}$ correspond to the ultraviolet and infrared cutoff scales, respectively. In order to obtain the leading-order scaling in $T$ of $\chi^{(2,1)}_{J_z P_z}(T)$, we can neglect the temperature independent term in Eq.~\eqref{chiJP1-2}. Performing the trace in Eq.~\eqref{chiJP1}, we obtain:
\begin{align}
\label{chiJP1-3}
&\chi^{(1)}_{J_z P_z}(T)
\sim \left(\frac{2\,e^2\, \Lambda_0}
{15\,\pi^2\, c\, T}\right)
T\,\sum \limits_{k_0}
\int \frac{d^3 {\mathbf k}}{(2\pi)^3}\, k_z \,
\frac{ 
 4\left (\mathrm{i}\,k_0-\frac{\mathbf{k}^2}{2m'}  \right )^3
\left( \partial_{k_z}\mathbf{d_k}\cdot \mathbf{d_k} \right)+12\left (\mathrm{i}\,k_0-\frac{\mathbf{k}^2}{2m'}  \right )
\left( \partial_{k_z}\mathbf{d_k}\cdot \mathbf{d_k}\right) |\mathbf{d_k}|^2 }
{\left[\left (\mathrm{i}\,k_0-\frac{\mathbf{k}^2}{2m'}  \right )^2-|\mathbf{d_k}|^2|\right]^3}.
\end{align}

For the two-loop diagram with the vertex correction (similar to Fig.~\ref{fig4}), we get the expression:
\begin{align}
\label{chiJP2}
& \chi^{(2,2)}_{J_z P_z}(T) = - T\, 
\sum \limits_{k_0}
\int \frac{d^3 {\mathbf k}}{(2\,\pi)^3}\,  
k_z \,\text{Tr}
\left[ \left \lbrace  \partial_{k_z}\mathbf{d}{(\mathbf k)}\cdot \mathbf \Gamma \right \rbrace
G_0(k_0,\mathbf{k})\, 
\tilde \Gamma_1(k_0,\mathbf{k})\, G_0(k_0,\mathbf{k}) \right],
\end{align}
where 
\begin{align}
\tilde \Gamma_1(k_0,\mathbf{k}) & =-\frac{2\,e^2}{c}\,T\,
\sum \limits_{\ell_0}\int \frac{d^3 {\boldsymbol \ell}}{(2\pi)^3}
\frac{G_0(k_0+\ell_0,\mathbf k+\boldsymbol \ell)\,
G_0(k_0+\ell_0,\mathbf k+\boldsymbol \ell)
}
{\boldsymbol \ell^2}
%%%%%%
=-\frac{e^2}{16\pi^2 c\, T}\left(\Lambda_0+\frac{2\,m}{3\,\Lambda_{IR}}
\,\big |\mathbf{d}{(\mathbf k)} \big |\right).
\label{chiJP2-2}
\end{align}
Plugging this in, we get: 
\begin{align}
\label{chiJP2-3}
&\chi^{(2,2)}_{J_z P_z}(T)
= \left(\frac{e^2}{16\pi^2 c\, T}\right)
T\,\sum \limits_{k_0}
\int \frac{d^3 {\mathbf k}}{(2\pi)^3}\, k_z
\left(\Lambda_0+\frac{2\,m}{3\,\Lambda_{IR}}|\mathbf{d}{(\mathbf k)}|\right)\frac{8\left (\mathrm{i}\,k_0-\frac{\mathbf{k}^2}{2m'}  \right )
\left( \partial_{k_z}\mathbf{d_k}\cdot \mathbf{d_k} \right)}
{\left[ \left (\mathrm{i}\,k_0-\frac{\mathbf{k}^2}{2m'}  \right )^2
-\big |\mathbf{d_k} \big |^2 \right]^2}\,.
\end{align}
In order to obtain the leading-order dependence on $T$, we can neglect the second term in Eq~\eqref{chiJP2-2}.

In Fig.~\ref{Fig:chi_JP_2loop}, we show the numerical result for $\chi^{2loop}_{J_zP_z}(T)=\chi^{(2,1)}_{J_z P_z}(T)+\chi^{(2,2)}_{J_z P_z}(T)$ as a function of temperature.

%%%%%%%%%%%%%%%%%%%%

\end{widetext}

 %%%%%%%%%%%%%%%%%%%%%%%%%%%%%%%%%%%
\end{document}